\documentclass[manuscript=article]{achemso}

\usepackage{siunitx}
\usepackage{booktabs}
\usepackage{multirow}
\usepackage{caption}
\usepackage{subcaption}
\usepackage[frozencache=true,cachedir=minted-cache]{minted}
\usepackage{amsmath}
\usepackage{textgreek}
\usepackage{hyperref}
\usepackage{fontenc}
\SectionNumbersOn
\DeclareUnicodeCharacter{207B}{$^-$}
\DeclareUnicodeCharacter{2202}{$\partial$}
\title{Clapeyron.jl: An extensible, open-source fluid-thermodynamics toolkit}

\author{Pierre J. Walker}
\affiliation{Equally contributing authors}
\alsoaffiliation[California Institute of Technology]
{Division of Chemistry and Chemical Engineering, California Institute of Technology, Pasadena, CA, 91125, U.S.A}
\alsoaffiliation[ICL]{Department of Chemical Engineering, Imperial College London, SW7 2AZ, United Kingdom}
\email{pjwalker@caltech.edu}
\author{Hon-Wa Yew}
\affiliation{Equally contributing authors}
\email{yewhonwa@gmail.com}
\author{Andrés Riedemann}
\affiliation[Concepcion]
{Department of Chemical Engineering, University of Concepción, Concepción, Región del Bio Bio, Chile}
\alsoaffiliation{Equally contributing authors}
\keywords{Equations of State, Cubics, SAFT, Activity, COSMO-SAC, Open-source, Thermodynamics, Julia}

\begin{tocentry}
\begin{center}
    \includegraphics[width=\textwidth]{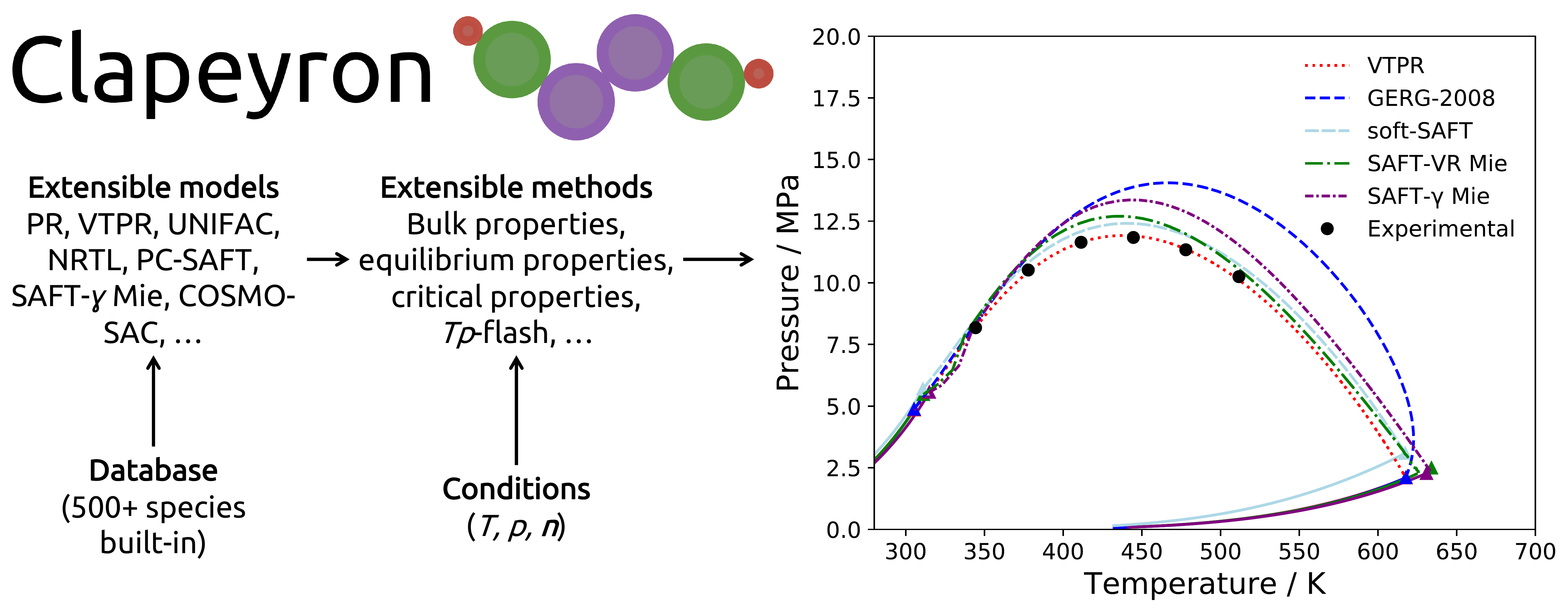}
\end{center}

\end{tocentry}

\begin{document}
\begin{abstract}
Thermodynamic models are often vital when characterising complex systems, particularly natural gas, electrolyte, polymer, pharmaceutical and biological systems. However, their implementations have historically been abstruse and cumbersome, and as such, the only options available were black-box commercial tools. In this article, we present \texttt{Clapeyron.jl}: a pioneering attempt at an open-source fluid-thermodynamics toolkit to build and make use of thermodynamic models. This toolkit is built in Julia, a modern language for scientific computing known for its ease of use, extensibility, and first-class support for differentiable programming. We currently support more models than any package available, including standard cubic (SRK, PR, PSRK, etc.), activity-coefficient (NRTL, UNIFAC, etc.), COSMO-based, and the venerable SAFT equations. The property-estimation methods supported are extensive, including bulk, VLE, LLE, VLLE and critical properties. With \texttt{Clapeyron.jl}, researchers and enthusiasts alike will be able to focus on the application and worry less about the implementation.
% Please include a maximum of seven keywords
\end{abstract}

\section{Introduction}
% 1. Context: where are equations of state used
% 2. Start from the beginning: Clapeyron and the ideal gas law
% 3. Build up to the new equations of state
% 4. Numerical methods
% 5. Current packages: commercial and open-source
% 6. In comes Clapeyron and Julia (and auto diff)
% 7. Layout

Thermodynamic models, such as equations of state and activity-coefficient models, are used to predict the properties of bulk matter at specified conditions. Today, they have found applications in natural gas\citep{Hendriks2010IndustrialProperties}, electrolyte\citep{Muller2020CalculationSystems,Walker2022ImportanceState}, polymer\citep{Gross2002ApplicationSystems,Tihic2008,Inguva2021Continuum-scaleThermodynamics}, metallic\citep{Walsh1957Shock-waveMetals}, pharmaceutical\citep{Hutacharoen2017PredictingPharmaceuticals,Mahmoudabadi2021ACompounds} and biological\citep{Cameretti2008ModelingPC-SAFT} systems. They have become ubiquitous in modelling for both academia and industry\citep{Hendriks2010IndustrialProperties} and, moreover, as the needs and demands placed upon this modelling grow ever more larger, these models have become more sophisticated. The thermodynamic modelling scene has now become quite complex, and it has become very difficult to keep track of all developments in the various fields.

\begin{figure}[h!]
    \centering
    \includegraphics[width=0.5\textwidth]{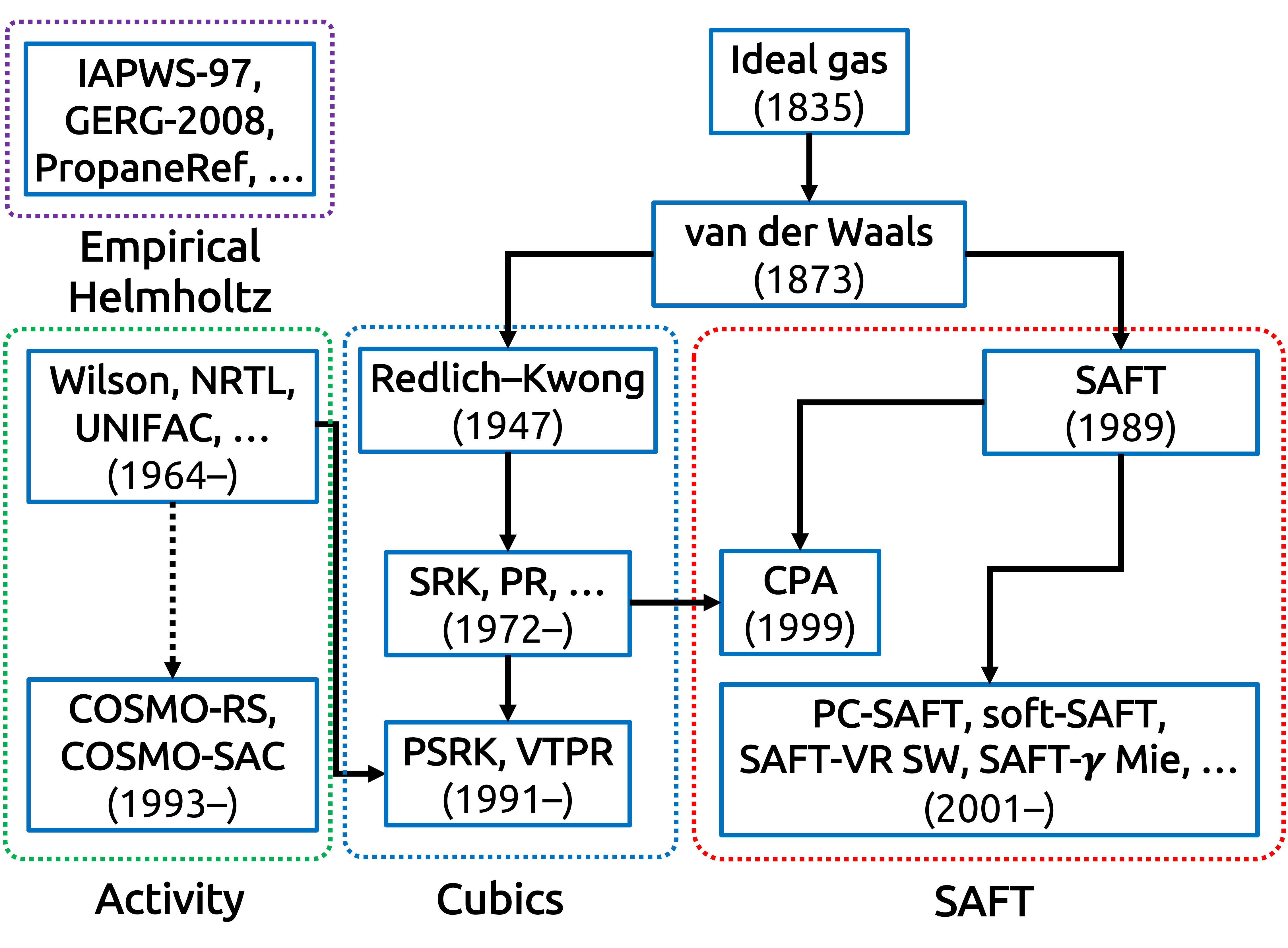}
    \caption{Condensed family tree of the thermodynamic models developed in the past two centuries.}
    \label{fig:timeline}
\end{figure}

The family tree of thermodynamic models, illustrated in figure \ref{fig:timeline}, begins in 1834, when \citet{Clapeyron1834MemoireChaleur} combined the laws of Boyle, Charles, Avogadro and Gay-Lussac, resulting in what we now refer to as the ideal gas law, which is arguably the most famous equation of state; indeed, it has become common knowledge to almost anyone with scientific literacy. It would not be until decades later that \citet{vanderWaals1873OverVloeistoftoestand} then introduced his equation of state---which paved the way to an entire class of equations of state known as the cubics. This class of equations proved to be extremely popular, primarily due to their simple functional form. Some of the notable cubic equations of state: Redlich--Kwong\citep{Redlich1949OnSolutions.} (RK), Soave--Redlich--Kwong\citep{Soave1972EquilibriumState} (SRK) and Peng--Robinson\citep{Peng1976AState} (PR), continue to be the most-used in industry today. Subsequent modifications of such equations have also been developed in the hopes of improving their accuracy whilst maintaining their simple functional form.\citep{Rackett1970EquationLiquids,Huron1979NewMixtures,Boston1980ProceedingsIndustries,Twu1980ImprovedBehavior,Peneloux1982AVolumes,Michelsen1990AState,Twu1992AState,Twu1995AEquation} These modifications alone are numerous and, for those unfamiliar with the field, can be difficult to navigate. 

The van der Waals' equation of state also turned out to be the foundation of another branch of models known as perturbation models, which are derived from first principles using statistical mechanics. It was not until the development of the Statistical Associating Fluid Theory (SAFT), introduced by \citet{Chapman1990NewLiquids}, that such equations became a viable alternative. SAFT-type equations retain a much-stronger physical basis than cubic equations, and include terms to explicitly account for dispersive and associative interactions. The use of SAFT-type equations is continually becoming more widespread; indeed, the popularity of the PC-SAFT equation, developed by \citet{Gross2001Perturbed-ChainMolecules}, rivals that of the cubic equations of state.  This increasing uptake of SAFT-type equations is a reflection of their broader applicability and more-predictive nature – but this flexibility has come at the cost of greater complexity in the equations which, together with their rapid development in recent decades, has meant that they have become very difficult to access for people outside the field.

In parallel to this tree, other types of thermodynamic models have also been developed. These include multi-parameter equations of state such as IAPWS-97\citep{Wagner2008InternationalIAPWS-IF97} and GERG-2008\citep{Kunz2012TheGERG-2004} which are restricted to specific systems but with levels of accuracy much higher than standard equations of state (we refer to them as Empirical Helmholtz equations of state). Another branch is the activity-coefficient models (such as the Wilson equation\citep{Wilson1964Vapor-LiquidMixing}, NRTL\citep{Renon1968LocalMixtures} and UNIFAC\citep{Fredenslund1977ComputerizedCoefficients}) which, although not as flexible as typical equations of state, have become very popular due to their simple form, large parameter databases and ease of computing typically complicated properties. Whilst also technically part of this branch, COSMO-based models\citep{Klamt1993COSMO:Gradient}, such as COSMO-RS\citep{Klamt1995Conductor-likePhenomena} and COSMO-SAC\citep{Lin2002AModel} have proven to be very powerful, fully predictive thermodynamic models. Activity-coefficient models have also been combined with cubic equations of state\citep{Huron1979NewMixtures,Michelsen1990AState} to improve the modelling of mixtures. This combination has resulted in some of the most-reliable equations of state, such as PSRK\citep{Holderbaum1991PSRK:UNIFAC} and VTPR\citep{Ahlers2001DevelopmentState}.

With such a wide selection of thermodynamic models, it is very easy to become overwhelmed. Unfortunately, the difficulty with these models does not end here. Typically, thermodynamic models provide a single property (usually the pressure, Helmholtz free energy or activity coefficients) at specified conditions (usually volume, temperature and composition). Obtaining other useful properties can be an arduous process. For example, just trying to obtain bulk properties such as the heat capacity will usually involve taking multiple derivatives of sometimes very complicated functions. As another example, obtaining vapour--liquid equilibrium (VLE) or critical properties will usually involve solving non-linear systems of equations made up of these high-order derivatives. Finally, flash routines, which solve for the phase equilibrium at specified conditions, making them one of the most important tools when using thermodynamic models, involve complicated, atypical optimisation problems; indeed, such is the complexity of the problem that an entire book \citep{Michelsen2007ThermodynamicAspects} has been written on the subject. 

With all these complicated aspects when using thermodynamic models, it shouldn't be surprising that, until recently, only commercial packages such as ASPEN\citep{2016AspenPlus} and gPROMS\citep{2020GPROMS} have provided the necessary tools to work with these models. The difficulty with such packages, aside from the financial barrier, is their `black-box' nature where, having access only to the front-end, users cannot modify the underlying code. Users are also limited to the selection of models provided by the package.

However, in recent years, the push for open-source software has led to the creation of many open-source thermodynamic modelling packages. Examples of this include \textit{CoolProp}\citep{Bell2014PureCoolProp} and \texttt{thermo}\citep{BellThermo:ChEDL}, although these place less emphasis on equations of state and activity-coefficient models. For packages focusing specifically on the latter, very recent examples include \texttt{phasepy}\citep{Chaparro2020Phasepy:Computation} written in Python, \texttt{thermopack}\citep{Wilhelmsen2017ThermodynamicMethods} written in FORTRAN, \texttt{FeOs}\citep{Rehner2022FeOsTheory} written in RUST and \texttt{teqp}\citep{Bell0ImplementingTeqp} written in C++. We note also that, famously, the PC-SAFT equation was released open-source when it was initially published, something that has unfortunately not been repeated for subsequent SAFT equations. 

Unfortunately, despite giving users complete access to source code, these endeavours still suffer from a few problems. Firstly, like their commercial counter-parts, they do not provide a wide range of thermodynamic models. Of course, it should be possible for users to use the open-source code as a template to implement their own models. However, as mentioned previously, in order to use these models in a productive way, one requires access to higher-order derivatives which will need to be obtained analytically in the case of \texttt{phasepy} and \texttt{thermopack}. To see where the difficulty arises here, it is recommended that readers look at \texttt{thermopack}'s documentation, which contains these derivatives. Particularly for SAFT-type models, these derivatives become very tedious. 

In the case of \texttt{FeOs} and \texttt{teqp}, tools such as hyper dual numbers\citep{Rehner2021ApplicationModeling} and automatic differentiation\citep{Leal2018AutodiffDifferentiation} are leveraged to avoid having to obtain derivatives analytically. The difficulty with these packages is linked to the programming language used; low-level languages such as Rust and C++ are chosen to maximise performance, but they are less user friendly. To make it accessible, an interpreted front-end (such as Python) is generally used, which helps with the adoption, but makes it difficult for a typical user to make extensions to the framework without substantial effort.
%(this is often referred to as the `two-language' problem). It therefore requires users to be familiar with at least two coding languages.

In this article we introduce \texttt{Clapeyron.jl}, a package named in honour of the progenitor of equations of state, that has been designed to avoid these difficulties. Like \texttt{FeOs} and \texttt{teqp}, \texttt{Clapeyron.jl} makes use of automatic differentiation but, unlike these packages, is written entirely in the novel, easy-to-use scientific-computing language, Julia. It already incorporates over 30 thermodynamic models, including all the well-known SAFT-type equations, many cubic equations and, in addition, activity-coefficient models and COSMO-SAC. Moreover, \texttt{Clapeyron.jl} is presented not just as open-source, but also as a user-extensible package, with users encouraged to submit their own tools and models for incorporation to the package, so that \texttt{Clapeyron.jl} can grow and become more flexible as the field develops.

Note that all figures in this manuscript can be re-created using the jupyter notebook file provided in the supplementary information. Whilst these figures contain multiple thermodynamic models, it is not the intention of this article to compare the performance of these models, only to illustrate specific functionalities of \texttt{Clapeyron.jl}.

The remainder of the article is set out as follows: in section \ref{sect:arch}, we provide details on \texttt{Clapeyron.jl}'s architecture, particularly model objects and functions that are the center of all computations. In section \ref{sect:eos}, we given a comprehensive overview of the thermodynamic models available in \texttt{Clapeyron.jl} and some of the special features available in each of them. Subsequently, in section \ref{sect:methods}, we describe how the various property estimation methods are implemented in an extensible way. We then highlight some current developments and future plans for the package in section \ref{sect:fw}. Finally, in section \ref{sect:conc}, we summarise the features of \texttt{Clapeyron.jl}. 

\section{Clapeyron.jl Architecture}
\label{sect:arch}
\begin{figure}[h!]
    \centering
    \includegraphics[width=0.5\textwidth]{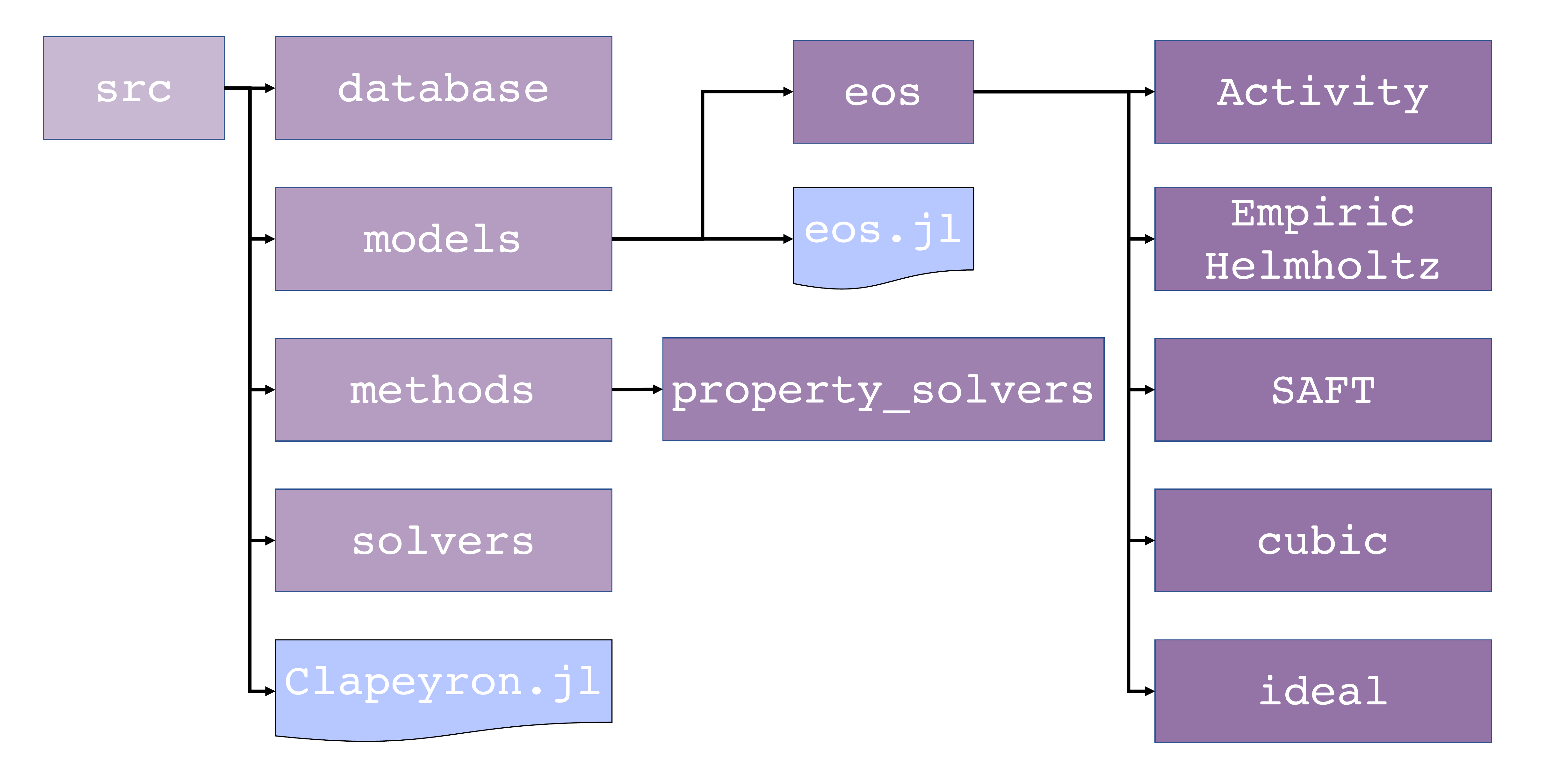}
    \caption{Architecture of the \texttt{Clapeyron.jl} package.}
    \label{fig:arch}
\end{figure}
\texttt{Clapeyron.jl} is divided into four main components, illustrated visually in figure \ref{fig:arch}:

\begin{itemize}
    \item \textbf{database} - the logic for extraction of parameters from comma-separated value files for the implementation into the \texttt{model} objects. More information on the structure of our databases can be fund in the supplementary information.
    \item \textbf{models} - functions that produce the \texttt{model} object that is at the center of \texttt{Clapeyron.jl}.
    \item \textbf{methods} - the thermodynamic methods to obtain thermodynamic properties from the models.    
    \item \textbf{solvers} - custom solvers provided by \texttt{Clapeyron.jl}.
\end{itemize}

This structure is designed to be easy to follow for brand-new users of the package. One can examine each aspect of \texttt{Clapeyron.jl} on the GitHub repository (see section \ref{sect:softav}). However, to appreciate how \texttt{Clapeyron.jl} works, one needs understand two of its key aspects: the \texttt{model} object and the \texttt{eos} function. These are discussed in detail in the subsequent sections.

\begin{figure}[h!]
    \centering
    \includegraphics[width=0.5\textwidth]{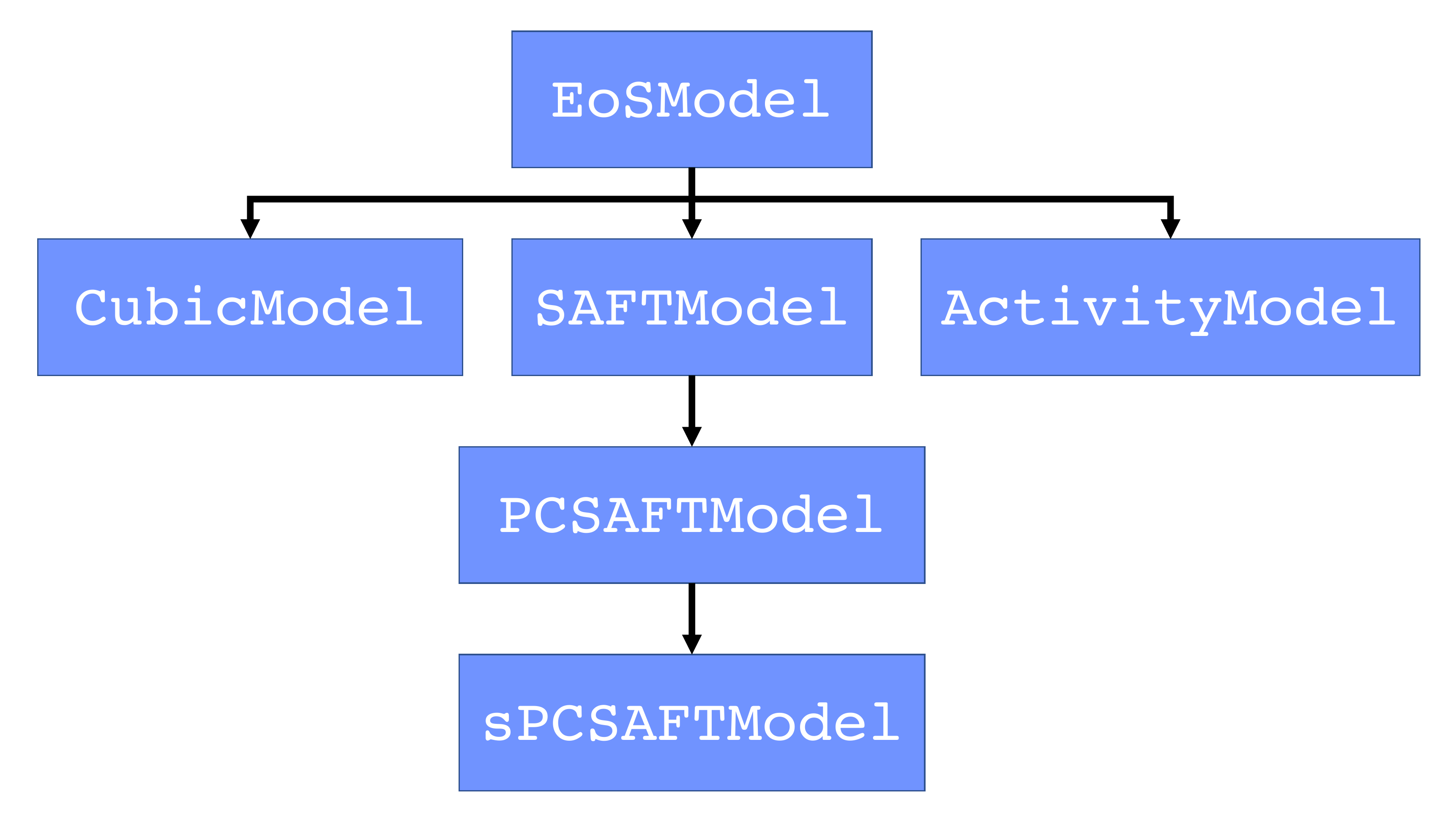}
    \caption{An example of the abstract type hierarchy in the \texttt{Clapeyron.jl}.}
    \label{fig:types}
\end{figure}
\subsection{Model object}
In \texttt{Clapeyron.jl}, thermodynamic models are created as structs (a composite type), which contain all of the parameters needed to completely identify the species and derive the thermodynamic properties. These structs are created as sub-types of an abstract type in order to create a hierarchy in which the model objects reside. Functions that are related to the models are written in the corresponding Julia script that dispatch on instances of these abstract types.

As an example, if we wanted to use the PC-SAFT equation of state for a water + carbon dioxide system, we would construct the \texttt{model} object by calling the constructor as:
\begin{minted}[breaklines,escapeinside=||,mathescape=true,  numbersep=1pt, gobble=2, frame=lines, fontsize=\small, framesep=2mm]{julia}
julia> model = PCSAFT(["water","carbon dioxide"])
PCSAFT{BasicIdeal} with 2 components:
 "water"
 "carbon dioxide"
Contains parameters: Mw, segment, sigma, epsilon, epsilon_assoc, bondvol

julia> model.params.epsilon
PairParam{PairParam{Float64}}["water", "carbon dioxide"]) with values:
2×2 Matrix{Float64}:
 366.51   250.508
 250.508  169.21

julia> model.references
2-element Vector{String}:
 "10.1021/ie0003887"
 "10.1021/ie010954d"
\end{minted}
As we can see above, the \texttt{model} object contains all the parameters that characterise the water+carbon dioxide system. We also include the references to the source of the parameters and thermodynamic model, ensuring that the developers of these equations receive the appropriate credit.

This \texttt{model} also has a type associated with it. In \texttt{Clapeyron.jl}, we have developed a hierarchy of types (a condensed version is illustrated in figure \ref{fig:types}), the head of which is the \texttt{EoSModel} abstract type. From this parent type, we divide the hierarchy into \texttt{SAFTModel}, \texttt{CubicModel}, \texttt{ActivityModel}, and so forth. The \texttt{SAFTModel} type is then further divided into types associated with each SAFT-type equation of state. This \texttt{model} object is a sub-type of the \texttt{PCSAFTModel} type. What this effectively means is that Julia will only perform operations on \texttt{model} using the functions whose input most closely resembles its type. 

To illustrate this we consider another \texttt{model} object constructed from the \texttt{sPCSAFT} function (i.e. \texttt{model2=sPCSAFT(["water","carbon dioxide"])}) and that, accordingly, will be associated with the \texttt{sPCSAFTModel} type (which is a sub-type of \texttt{PCSAFTModel}). Only three equations are different between the theories expressed in the PC-SAFT and sPC-SAFT equations of state. One of these is the reduced Helmholtz free energy for a hard-sphere fluid. Within \texttt{Clapeyron.jl}, the function corresponding to this equation is \texttt{a\_hs}, written for each equation of state as:
\begin{minted}[breaklines,escapeinside=||,mathescape=true,  numbersep=1pt, gobble=2, frame=lines, fontsize=\small, framesep=2mm]{julia}

# PC-SAFT code
function a_hs(model::PCSAFTModel, V, T, z,_data=@f(data))
    _,ζ0,ζ1,ζ2,ζ3,_ = _data
    return 1/ζ0 * (3ζ1*ζ2/(1-ζ3) + ζ2^3/(ζ3*(1-ζ3)^2) + (ζ2^3/ζ3^2-ζ0)*log(1-ζ3))
end

# sPC-SAFT code
function a_hs(model::sPCSAFTModel, V, T, z,_data=@f(data))
    _,_,_,_,η,_ = _data
    return (4η-3η^2)/(1-η)^2
end
    \end{minted}
Calling \texttt{a\_hs} using the \texttt{model} object will tell Julia to only use the first function, whilst calling \texttt{a\_hs} using the \texttt{model2} will tell Julia to use the latter function. This powerful feature allows developers to define variants of thermodynamic models (like sPC-SAFT) with very few lines of code, and even define unique methods for specific models, should the need arise.

The above is also an example of how \texttt{Clapeyron.jl} makes full use of the Unicode characters, bringing the code much closer to how the corresponding equations are presented in their original publications. The use of Unicode characters is strongly supported within the Julia community.

\subsubsection{Group contribution}
\texttt{Clapeyron.jl} also supports group-contribution (GC) models, where parameters are associated with groups which are combined to represent species. Some GC models that we currently support include SAFT-$\gamma$ Mie\citep{Papaioannou2014GroupSegments} equation of state, the UNIFAC\citep{Weidlich1987A.gamma..infin.} activity-coefficient model and, Walker's model\citep{Walker2020AState} and the Joback Method\citep{JOBACK1987ESTIMATIONGROUP-CONTRIBUTIONS}  for ideal-gas heat capacities. In order to form species from groups within \texttt{Clapeyron.jl}, one can specify the type and number of groups within a species as using a tuple containing the name of the species, and a list of key-value pairs where the key is the name of the group and the value is the group multiplicity:
\begin{minted}[breaklines,escapeinside=||,mathescape=true,  numbersep=1pt, gobble=2, frame=lines, fontsize=\small, framesep=2mm]{julia}
julia> model = SAFTgammaMie([("ethanol",["CH3"=>3,"CH2OH"=>1]),("ethyl acetate", ["CH3"=>2,"COO"=>1,"CH2"=>1])])
SAFTgammaMie{BasicIdeal} with 2 components:
"ethanol": "CH3"=>3, "CH2OH"=>1
"ethyl acetate": "CH3"=>2, "COO"=>1, "CH2"=>1
Contains parameters: segment, mixedsegment, shapefactor, lambda_a, lambda_r, sigma, epsilon, epsilon_assoc, bondvol
\end{minted}
This is in a similar vein to how species are constructed from groups in tools such as gPROMS\citep{2020GPROMS}. 

\subsection{\texttt{eos} function}
With a large number of thermodynamic models available within a single package, it is necessary to ensure that all models have a common structure at some level. In \texttt{Clapeyron.jl}, we dispatch on a function called \texttt{eos}, which is the Helmholtz free energy. This function is at the center of all computations in \texttt{Clapeyron.jl}. For most models, we would generally make use of the built-in function that dispatches on the top-level \texttt{EoSModel}, where it is made up of an ideal and residual contribution (since most thermodynamic models are built this way):
\begin{minted}[breaklines,escapeinside=||,mathescape=true,  numbersep=1pt, gobble=2, frame=lines, fontsize=\small, framesep=2mm]{julia}
# General function
function eos(model::EoSModel, V, T, z=SA[1.0])
    return N_A*k_B*Σ(z)*T * (a_ideal(idealmodel(model),V,T,z)
        +a_res(model,V,T,z))
end

# Ideal models only
function eos(model::IdealModel, V, T, z=SA[1.0])
    return N_A*k_B*Σ(z)*T *a_ideal(model,V,T,z)
end
\end{minted}
Here \texttt{V} is the total volume (in m$^3$), \texttt{T} is the temperature (in K) and \texttt{z} is the amount of components (in mol).

% When creating an equation of state in \texttt{Clapeyron.jl}, users would generally dispatch on the \texttt{a\_res(model,V,T,z)} function for the reduced Helmholtz free energy, while still using one of \texttt{Clapeyron.jl}'s built-in ideal models (Users could equally build custom ideal models to work with existing residual models if they so wish).

It is, of course, possible to express the Helmholtz free energy in a different way. For example, the if \texttt{model} is a purely ideal model, there is no residual contribution. Using multiple dispatch, we can define an \texttt{eos} function specific to ideal models (using the \texttt{IdealModel} type).

% \subsection{Methods}
% Methods are where the equations of state defined in the models are actually used to derive thermodynamic properties.

% Clapeyron.jl uses \texttt{ForwardDiff.jl} to obtain the derivatives of the Helmholtz free energy with respect to the temperature, volume and the compositions. Several helper functions are defined to abstract the procedure.

% code here

% \subsection{Solvers}
% Clapeyron.jl provides a subpackage for generic custom-made solvers that are useful for thermodynamic modelling applications. These can be imported independently of the base Clapeyron.jl package if one were to find a suitable application for it outside of this package.

%%%%%%%%%%%%%%%
%Numerous features are available within Clapeyron.jl, both within the base package and beyond. We shall first provide information on the former before showing examples as to how the package can be extended beyond the base features.
% At the outset, both equations of state and methods required to use them are available within Clapeyron.jl. We shall first go over the models available, pointing out some of the special features for each of them, before providing information of how Clapeyron.jl uses unique methods to provide a variety of properties for all models along with how the Julia language has enabled this. 
\section{Models}
\label{sect:eos}
\begin{table*}
\caption{List of supported thermodynamic models within \texttt{Clapeyron.jl}, along with their function names. Models that are sub-types of a main model that are also supported are indicated as variants.}
\label{tbl:models}
\resizebox{\textwidth}{!}{
\begin{tabular}{lllll}
\textbf{Type}                                & \textbf{Main}                           & \textbf{Function}                  & \textbf{Variant}              & \textbf{Function}    \\ \hline
\multirow{6}{*}{Cubics}             & van der Waals\citep{vanderWaals1873OverVloeistoftoestand}                  & \texttt{vdW}                       &                      &             \\ \cline{2-5} 
                                    & \multirow{2}{*}{Redlich-Kwong\citep{Redlich1949OnSolutions.}} & \multirow{2}{*}{\texttt{RK}}       & Soave-Redlich-Kwong\citep{Soave1972EquilibriumState}  & \texttt{SRK}         \\ \cline{4-5} 
                                    &                                &                           & Predictive SRK\citep{Horstmann2005PSRKComponents}       & \texttt{PSRK}        \\ \cline{2-5} 
                                    & \multirow{3}{*}{Peng-Robinson\citep{Peng1976AState}} & \multirow{3}{*}{\texttt{PR}}       & PR-78\citep{Robinson1978ThePrograms}                & \texttt{PR78}        \\ \cline{4-5} 
                                    &                                &                           & UMR-PR\citep{Voutsas2004UniversalState}               & \texttt{UMRPR}       \\ \cline{4-5} 
                                    &                                &                           & Volume-Translated PR\citep{Ahlers2001DevelopmentState} & \texttt{VTPR}        \\ \hline
              & SAFT\citep{chapman89,Chapman1990NewLiquids}                           & \texttt{ogSAFT}                    &                      &             \\ \cline{2-5} 
                                    & CK-SAFT\citep{Huang1990EquationMolecules,Huang1991EquationMixtures}                        & \texttt{CKSAFT}                    & simplified CK-SAFT\citep{Fu1995AMixtures}   & \texttt{sCKSAFT}     \\ \cline{2-5} 
                                    & BACK-SAFT\citep{Chen2001EquationPoint}                      & \texttt{BACKSAFT}                  &                      &             \\ \cline{2-5} 
                                    & LJ-SAFT\citep{Kraska1996PhaseWater,Kraska1996PhaseWaterb}                        & \texttt{LJSAFT}                    &                      &             \\ \cline{2-5} 
                                    & soft-SAFT\citep{Blas1997ThermodynamicTheory}                      & \texttt{softSAFT}                  &                      &             \\ \cline{2-5} 
                                    & CPA\citep{Kontogeorgis1996AnFluids,Yakoumis1997Vapor-liquidState}                            & \texttt{CPA}                       & simplified CPA\citep{Kontogeorgis1999MulticomponentMixtures}       & \texttt{sCPA}        \\ \cline{2-5} 
                                    & SAFT-VR SW\citep{Gil-Villegas1997StatisticalRange}                     & \texttt{SAFTVRSW}                  &                      &             \\ \cline{2-5} 
                                    & PC-SAFT\citep{Gross2001Perturbed-ChainMolecules,Gross2002ApplicationSystems}                        & \texttt{PCSAFT}                    & simplified PC-SAFT\citep{vonSolms2003}   & \texttt{sPCSAFT}     \\ \cline{2-5} 
                                    & SAFT-VR Mie\citep{Lafitte2013AccurateSegments,Dufal2015TheFluids}                    & \texttt{SAFTVRMie}                 & SAFT-VRQ Mie\citep{Aasen2019EquationDeuterium}         & \texttt{SAFTVRQMie}  \\ \cline{2-5} 
\multirow{-10}{*}{SAFT}                                    & SAFT-$\gamma$ Mie\citep{Papaioannou2014GroupSegments,Papaioannou2016ApplicationIndustry}              & \texttt{SAFTgammaMie}              &                      &             \\ \hline
\multirow{8}{*}{Activity}           & Wilson\citep{Wilson1964Vapor-LiquidMixing}                         & \texttt{Wilson}                    &                      &             \\ \cline{2-5} 
                                    & NRTL\citep{Renon1968LocalMixtures}                           & \texttt{NRTL}                      &                      &             \\ \cline{2-5} 
                                    & UNIQUAC\citep{Abrams1975StatisticalSystems}                        & \texttt{UNIQUAC}                   &                      &             \\ \cline{2-5} 
                                    & \multirow{2}{*}{UNIFAC}                         & \multirow{2}{*}{N/A} & og-UNIFAC\citep{Fredenslund1977ComputerizedCoefficients}                                       &  \texttt{ogUNIFAC}           \\ \cline{4-5} 
                                    &                                &                           & mod-UNIFAC\citep{Weidlich1987A.gamma..infin.}        & \texttt{UNIFAC} \\ \cline{2-5} 
                                    & \multirow{3}{*}{COSMO-SAC}     & \multirow{3}{*}{N/A} & COSMO-SAC-02\citep{Lin2002AModel}         & \texttt{COSMOSAC02}  \\ \cline{4-5} 
                                    &                                &                           & COSMO-SAC-10\citep{Hsieh2010ImprovementsPredictions}         & \texttt{COSMOSAC10}  \\ \cline{4-5} 
                                    &                                &                           & COSMO-SAC-dsp\citep{Hsieh2014ConsideringBehavior}        & \texttt{COSMOSACdsp} \\ \hline
                                                  & Reid Polynomial\citep{Reid1959TheLiquids}                & \texttt{ReidIdeal}                 &                      &             \\ \cline{2-5} 
                                    & Joback Method\citep{JOBACK1987ESTIMATIONGROUP-CONTRIBUTIONS}                  & \texttt{JobackIdeal}               &                      &             \\ \cline{2-5} 
\multirow{-3}{*}{Ideal}                                    & Walker Model\citep{Walker2020AState}                   & \texttt{WalkerIdeal}               &                      &             \\ \hline

\multirow{3}{*}{Empirical Helmholtz} & IAPWS-95\citep{Wagner2008InternationalIAPWS-IF97}                       & \texttt{IAPWS95}                   &                      &             \\ \cline{2-5} 
                                    & GERG-2008\citep{Kunz2012TheGERG-2004}                      & \texttt{GERG2008}                  &                      &             \\ \cline{2-5} 
                                    & Propane Reference\citep{Lemmon2009ThermodynamicMPa}              & \texttt{PropaneRef}                &                      &             \\ \hline
SPUNG                               & SPUNG\citep{Jrstad1993EquationHydrocarbons}                          & \texttt{SPUNG}                     &                      &             \\ \hline
\end{tabular}}
\end{table*}
\texttt{Clapeyron.jl} supports over 30 benchmarked thermodynamic models, most of which are listed in table \ref{tbl:models}. The highly flexible nature of \texttt{Clapeyron.jl} enables us to express many thermodynamic models in a modular way, which allows users to mix-and-match different components in an unprecedented manner.

\subsection{Cubic equations of state}
\begin{table*}
\caption{List of supported $\alpha$-functions, volume translation and mixing rule methods wiht \texttt{Clapeyron.jl}, along with their function names.}
\label{tbl:cubic_mod}
\centering
\begin{tabular}{lll}
\textbf{Component}                           & \textbf{Name}                              & \textbf{Function}            \\\hline
\multirow{7}{*}{Alpha function}     & Redlich--Kwong\citep{Redlich1949OnSolutions.}                                & \texttt{RKAlpha}             \\\cline{2-3}
                                    & Soave\citep{Soave1972EquilibriumState}                             & \texttt{SoaveAlpha}          \\\cline{2-3}
                                    & Peng--Robinson\citep{Peng1976AState}                                & \texttt{PRAlpha}             \\\cline{2-3}
                                    & PR-78\citep{Robinson1978ThePrograms}                              & \texttt{PR78Alpha}           \\\cline{2-3}
                                    & Boston-Matthias\citep{Boston1980ProceedingsIndustries}                   & \texttt{BMAlpha}             \\\cline{2-3}
                                    & Twu\citep{Twu1992AState}                               & \texttt{TwuAlpha}            \\\cline{2-3}
                                    & Magoulas-Tassios\citep{Magoulas1990ThermophysicalOnes}                                & \texttt{MTAlpha}             \\\hline
\multirow{3}{*}{Volume Translation} & Rackett\citep{Rackett1970EquationLiquids}                           & \texttt{RackettTranslation}  \\\cline{2-3}
                                    & Peneloux\citep{Peneloux1982AVolumes}                          & \texttt{PenelouxTranslation} \\\cline{2-3}
                                    & Magoulas-Tassios\citep{Magoulas1990ThermophysicalOnes}                                & \texttt{MTTranslation}       \\\hline
\multirow{7}{*}{Mixing rule}        & van der Waals one-fluid\citep{Waals1890MolekulartheorieBesteht}           & \texttt{vdW1fRule}           \\\cline{2-3}
                                    & Kay\citep{Kontogeorgis2009ThermodynamicTheories}                               & \texttt{KayRule}             \\\cline{2-3}
                                    & Huron-Vidal\citep{Huron1979NewMixtures}                       & \texttt{HVRule}              \\\cline{2-3}
                                    & Modified Huron-Vidal first order\citep{Michelsen1990AState}  & \texttt{MHV1Rule}            \\\cline{2-3}
                                    & Modified Huron-Vidal second order\citep{Michelsen1990AState} & \texttt{MHV2Rule}            \\\cline{2-3}
                                    & LCVM\citep{Boukouvalas1994PredictionUNIF}                              & \texttt{LCVMRule}            \\\cline{2-3}
                                    & Wong-Sandler\citep{Wong1992AState}                      & \texttt{WSRule}   \\\hline          
\end{tabular}
\end{table*}
In general, cubics are written in the following form:
\begin{equation}
    p = \frac{Nk_\text{B}T}{V-Nb}+\frac{a(T)}{(V-Nb_1)(V-Nb_2)}\,,
    \label{eq:cubic}
\end{equation}
where $p$, $N$, $V$, $T$ and $k_\text{B}$ are the pressure, total number of particles, volume, temperature and the Boltzmann constant, respectively. $a(T)$ and $b$ are related to the inter-molecular interactions and excluded volume of the system, respectively. $b_1$ and $b_2$ are related to the parameter $b$ by a certain factor dependent on the equation of state.

\begin{figure}[h!]
\centering
  \begin{subfigure}[b]{0.49\textwidth}
    \includegraphics[width=1\textwidth]{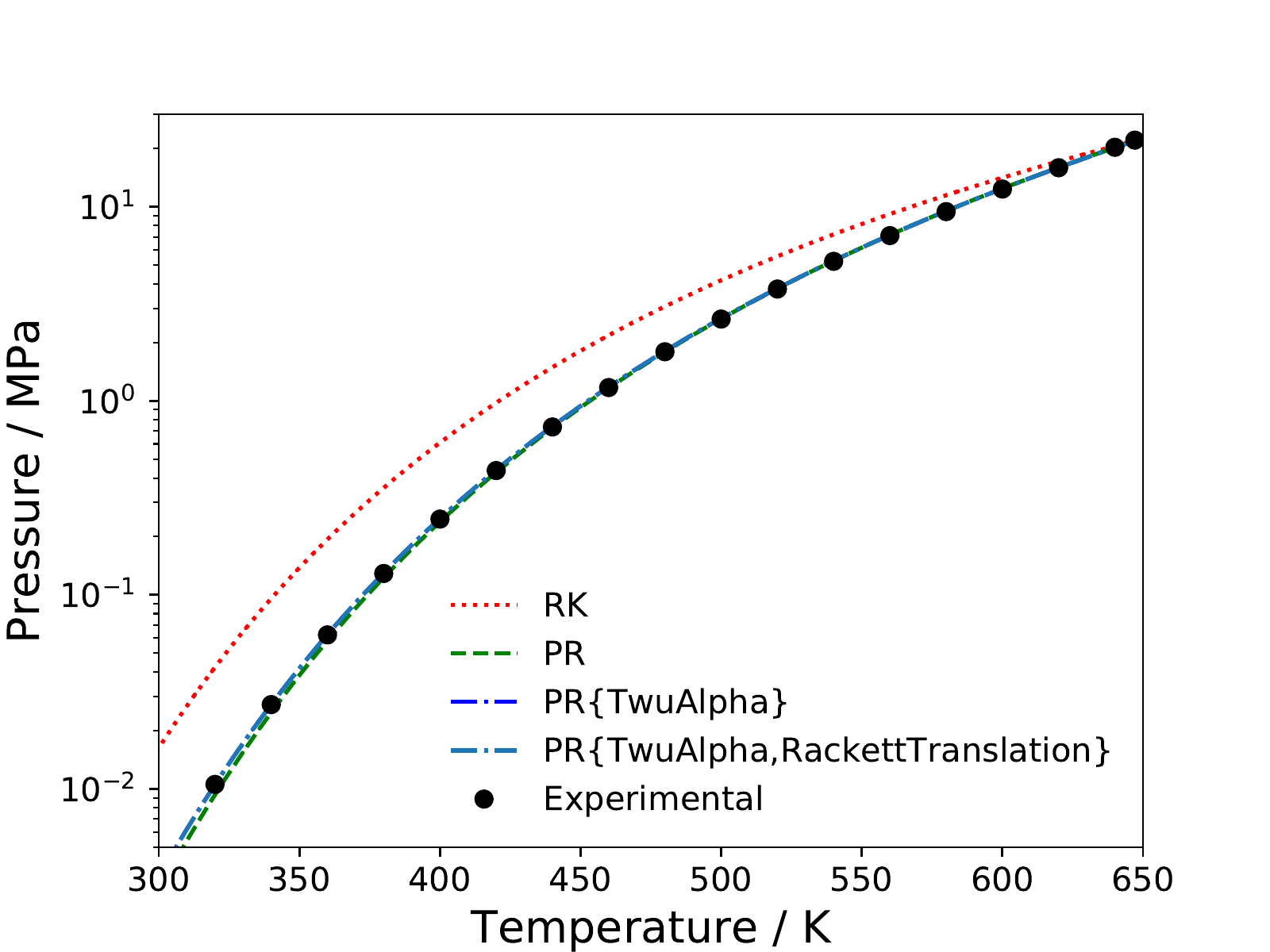}
    \caption{Saturation curve of water for different $\alpha$ functions and volume translations.\citep{Lemmon2020ThermophysicalSystems}}
    \label{fig:water_psat_cubic}
  \end{subfigure}
  \begin{subfigure}[b]{0.49\textwidth}
    \includegraphics[width=1\textwidth]{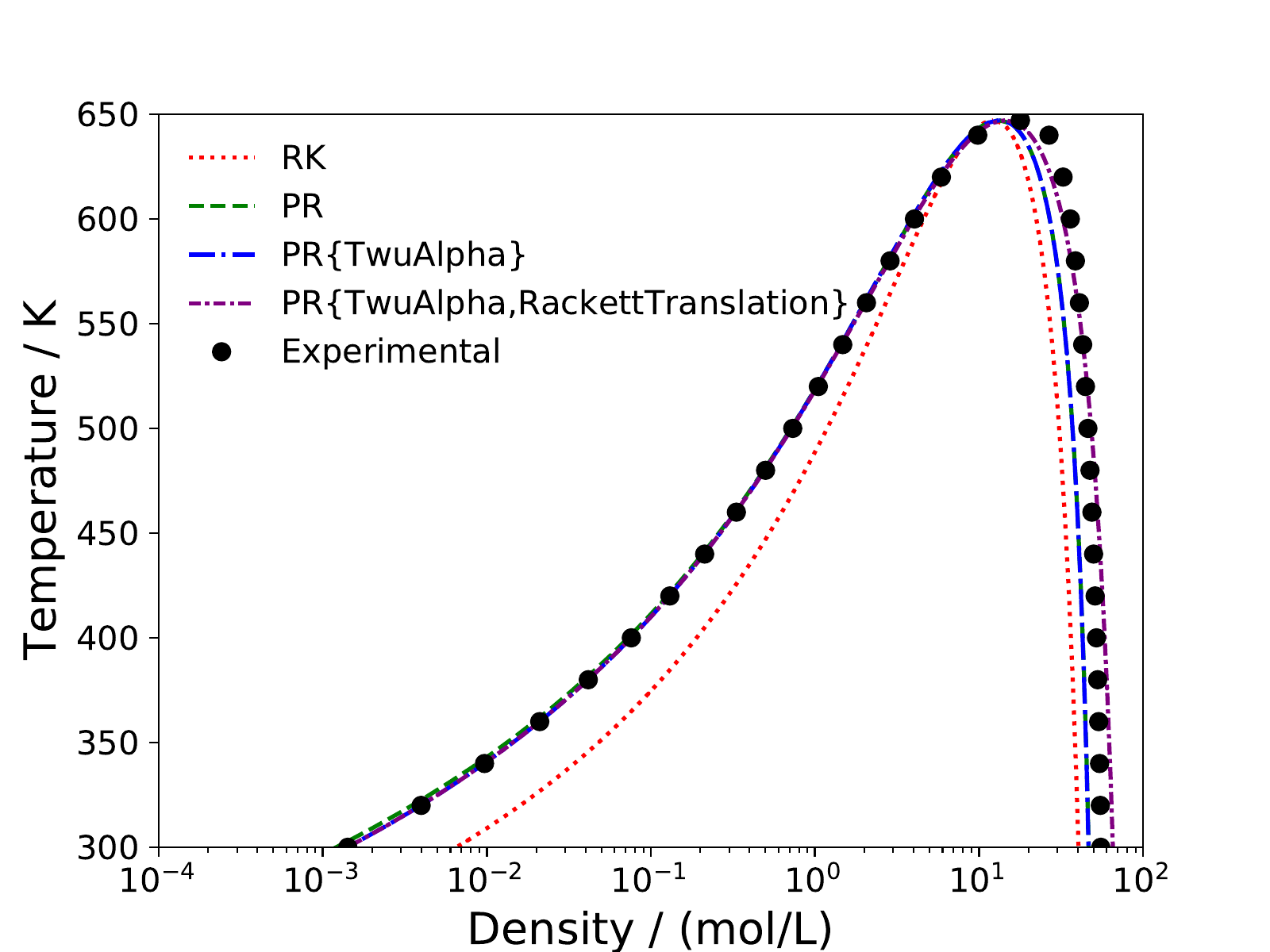}
    \caption{VLE envelope of water for different $\alpha$ functions and volume translations.\citep{Lemmon2020ThermophysicalSystems}}
    \label{fig:water_vle_cubic}
  \end{subfigure}
  \begin{subfigure}[b]{0.49\textwidth}
    \includegraphics[width=1\textwidth]{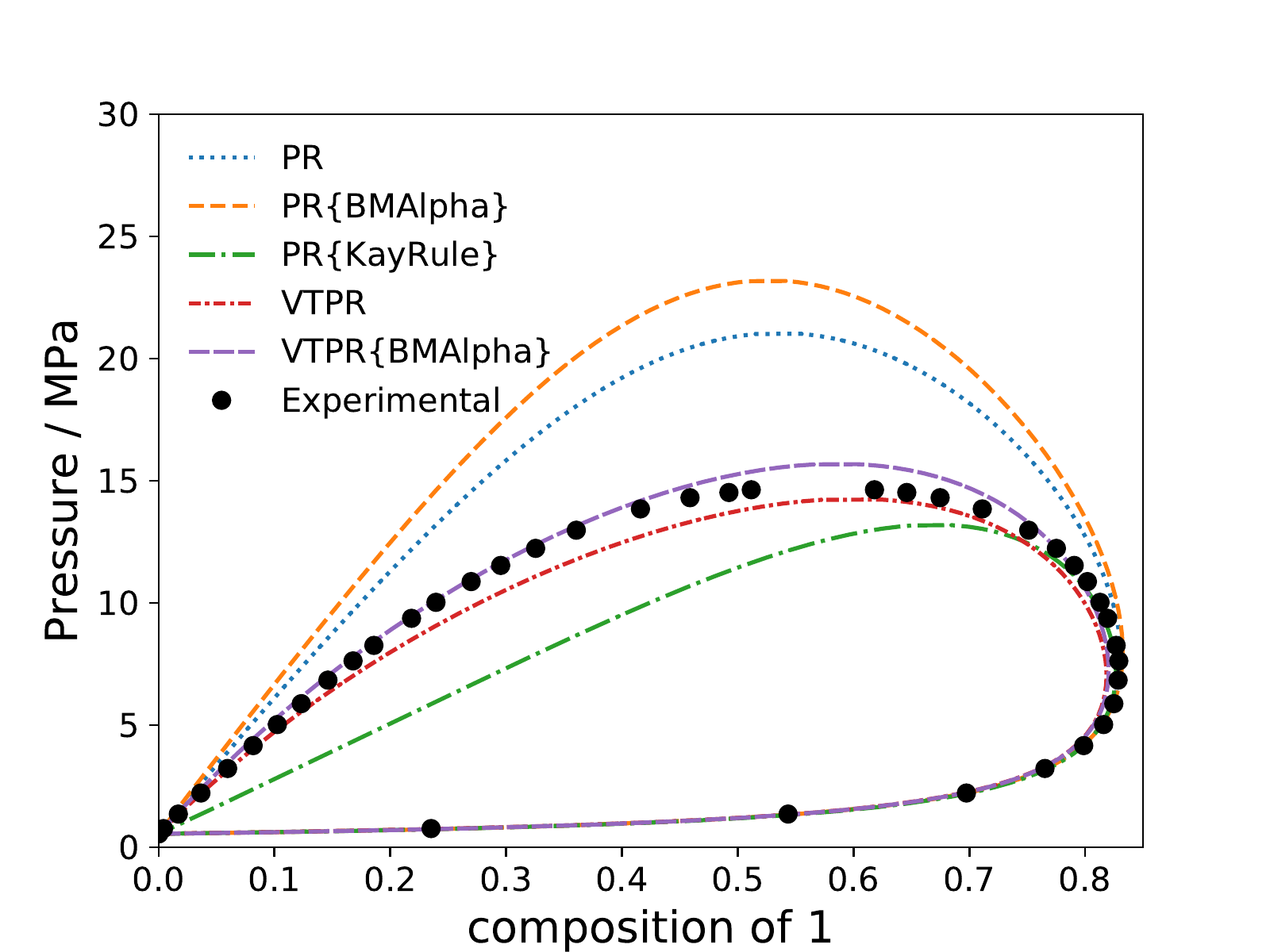}
    \caption{$pxy$ diagram of CO (1) + CO$_2$ (2)  at 218.15\,K for different $\alpha$ functions and mixing rules.\citep{Souza2018MeasurementMPa}}
    \label{fig:co2_co_alpha}
  \end{subfigure}
  \begin{subfigure}[b]{0.49\textwidth}
    \includegraphics[width=1\textwidth]{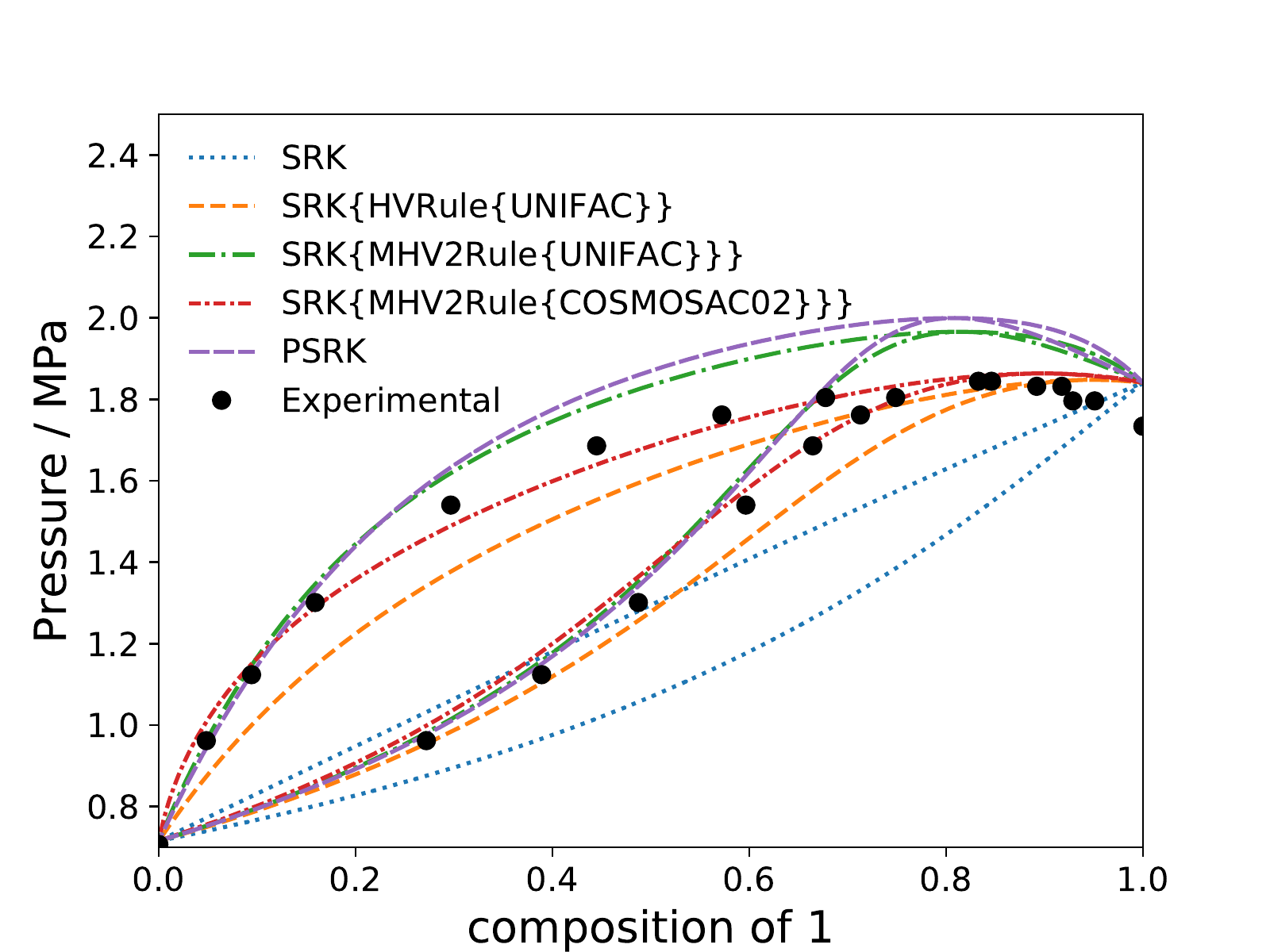}
    \caption{$pxy$ diagram of methanol (1) + benzene (2) at 433.15\,K for different mixing rules.\citep{Butcher2007ThermodynamicTemperatures}}
    \label{fig:meoh_benz_mix}
  \end{subfigure}
  \caption{Examples of the different capabilities of \texttt{Clapeyron.jl} for cubic equations of state.}
\end{figure}

In contrast to many of the other models listed in table \ref{tbl:models}, cubic equations of state offer a great deal of flexibility in three particular aspects (these have been summarised in table \ref{tbl:cubic_mod}):
\begin{itemize}
    \item $\alpha$-functions: Initially introduced by \citet{Soave1972EquilibriumState} to improve the predictions made by the Redlich--Kwong equation\citep{Redlich1949OnSolutions.}, numerous $\alpha$-functions have been developed to improve the performance of the equations of state. The $\alpha$-functions typically appear within the function $a(T)$ shown in equation \ref{eq:cubic} as:
    \begin{equation}
        a(T)=a\alpha(T)\,,
    \end{equation}
    where $a$ is a parameter usually related to the critical point of a species and $\alpha(T)$ is the $\alpha$-function. To give an example of this, the $\alpha$-function within the Redlich--Kwong equation is:
    \begin{equation}
        \alpha(T)=\frac{1}{\sqrt{T_r}}\,,
    \end{equation}
    where $T_r$ is the reduced temperature relative to the critical point. The Soave–Redlich–Kwong equation differs from the RK equation through its $\alpha$-function:
    \begin{equation}
        \alpha(T)=\left(1+(0.480+1.547\omega-0.176\omega^2)\left(1-\sqrt{T_r}\right)\right)^2\,,
    \end{equation}
    
    where $\omega$ is the acentric factor. Within \texttt{Clapeyron.jl}, it is possible to switch between $\alpha$-functions by simply providing an additional optional argument. This is, in fact, how we construct the Soave--Redlich--Kwong equation (rather than writing an entirely separate section of code):
\begin{minted}[breaklines,escapeinside=||,mathescape=true,  numbersep=1pt, gobble=2, frame=lines, fontsize=\small, framesep=2mm]{julia}
julia> model = SRK(["butane"])
RK{BasicIdeal, SoaveAlpha, NoTranslation, vdW1fRule} with 1 component:
 "butane"
Contains parameters: a, b, Tc, Pc, Mw

julia> model = SRK(["butane"];alpha=BMAlpha)
RK{BasicIdeal, BMAlpha, NoTranslation, vdW1fRule} with 1 component:
 "neon"
Contains parameters: a, b, Tc, Pc, Mw
\end{minted}

    It is equally as easy to switch to another $\alpha$-function listed in table \ref{tbl:cubic_mod}, as shown above.  Examples of this are shown in figures \ref{fig:water_psat_cubic} and \ref{fig:water_vle_cubic}, where switching $\alpha$-functions can improve the accuracy of our results. Furthermore, if one needs to model a species beyond its critical point, the Boston--Mathias $\alpha$-function can be specified, which can also have an impact on VLE properties as shown in figure \ref{fig:co2_co_alpha}.
    
    \item Volume translation: One often finds that the cubic equations perform poorly when predicting liquid densities. To improve upon this without affecting the simple functional form of the equation, one can `translate' the volume used in the equation of state by a certain value $c$:
    \begin{equation}
        V=V_\text{EoS}+Nc\,,
    \end{equation}
    where $V_\text{EoS}$ represents the untranslated volume in the equation of state. There are a few different methods which are listed in table \ref{tbl:cubic_mod} and, like the $\alpha$-functions, it is very easy to switch between them for a particular cubic equation:
    \begin{minted}[breaklines,escapeinside=||,mathescape=true,  numbersep=1pt, gobble=2, frame=lines, fontsize=\small, framesep=2mm]{julia}
julia> model = PR(["water"])
PR{BasicIdeal, PRAlpha, NoTranslation, vdW1fRule} with 1 component:
 "water"
Contains parameters: a, b, Tc, Pc, Mw
julia> volume(model,1e5,298.15)
2.1390770714552146e-5
julia> model = PR(["water"];
translation=RackettTranslation)
PR{BasicIdeal, PRAlpha, RackettTranslation, vdW1fRule} with 1 component:
 "water"
Contains parameters: a, b, Tc, Pc, Mw
julia> volume(model,1e5,298.15)
1.8705398679867126e-5
    \end{minted}
    We illustrate the capability of volume translation in \texttt{Clapeyron.jl} in figures \ref{fig:water_psat_cubic} and \ref{fig:water_vle_cubic}; as shown in these figures, whilst it doesn't have an impact on the saturation curve, it can improve the predicted liquid densities.
    
    \item Mixing rules: When one wishes to model mixtures with a cubic equation of state, mixing rules are required to obtain the parameters $a$ and $b$ in equation \ref{eq:cubic}. The most famous and simplest of these is the van der Waals one-fluid mixing rule:
    \begin{equation}
        \bar{a} = \sum_i\sum_jx_ix_ja_{ij}\,,
    \end{equation}
    \begin{equation}
        \bar{b} = \sum_i\sum_jx_ix_jb_{ij}\,,
    \end{equation}
    where $x_i$ denotes the composition of species $i$, the over bars indicate the parameters characterising the one-fluid average and subscripts $ij$ denote parameters characterising the interactions between species $i$ and $j$. When $i=j$ we refer to the familiar species-specific parameters; when $i\neq j$, we are referring to the unlike parameters which are typically obtained from combining rules:
    \begin{equation}
        a_{ij} = \sqrt{a_{ii}a_{jj}}(1-k_{ij})\,,
    \end{equation}
    \begin{equation}
        b_{ij} = \frac{b_{ii}+b_{jj}}{2}\,,
    \end{equation}
    where $k_{ij}$ is the binary interaction parameter between $i$ and $j$. The van der Waals one-fluid mixing rule provides satisfactory results when dealing with similar species. For more-complicated systems, advanced mixing rules should be used. Examples of these are listed within table \ref{tbl:cubic_mod}; details on the formulation of these can be found in their respective references. 
    What sets these mixing rules apart from the van der Waals one-fluid (and Kay) rule is that an activity coefficient model must be specified as well. The flexibility of \texttt{Clapeyron.jl} allows one to choose between any of the activity-coefficient models listed in table \ref{tbl:models}. For example:
\begin{minted}[breaklines,escapeinside=||,mathescape=true,  numbersep=1pt, gobble=2, frame=lines, fontsize=\small, framesep=2mm]{julia}
julia> model = PR(["methanol","benzene"];mixing=KayRule)
PR{BasicIdeal, PRAlpha, NoTranslation, KayRule} with 2 components:
 "methanol"
 "benzene"
Contains parameters: a, b, Tc, Pc, Mw
julia> model = PR(["methanol","benzene"];mixing=HVRule, activity=UNIFAC)
PR{BasicIdeal, PRAlpha, NoTranslation, HVRule{UNIFAC}} with 2 components:
 "methanol"
 "benzene"
Contains parameters: a, b, Tc, Pc, Mw
    \end{minted}
    One can observe, in figures \ref{fig:co2_co_alpha} and \ref{fig:meoh_benz_mix}, the impact of using different mixing rules on the predicted VLE behaviour of a mixture. 
\end{itemize}
As one can imagine, with the flexibility to switch between $\alpha$-functions, volume-translation methods and mixing rules with nothing more than a change in an input argument, \texttt{Clapeyron.jl} is able to provide a variety of cubic equations of state with a relatively small amount of code. An example of the benefits of this is the ability to `create' some of the high-accuracy cubic equations of state such as Predictive-SRK\citep{Horstmann2005PSRKComponents} (example calculations illustrated in figure \ref{fig:co2_co_alpha}) and Volume-Translated-PR\citep{Ahlers2001DevelopmentState} (exemplified in figure \ref{fig:meoh_benz_mix}).
\subsection{SAFT equations of state}
\begin{figure}[h!]
\centering
  \begin{subfigure}[b]{0.49\textwidth}
    \includegraphics[width=1\textwidth]{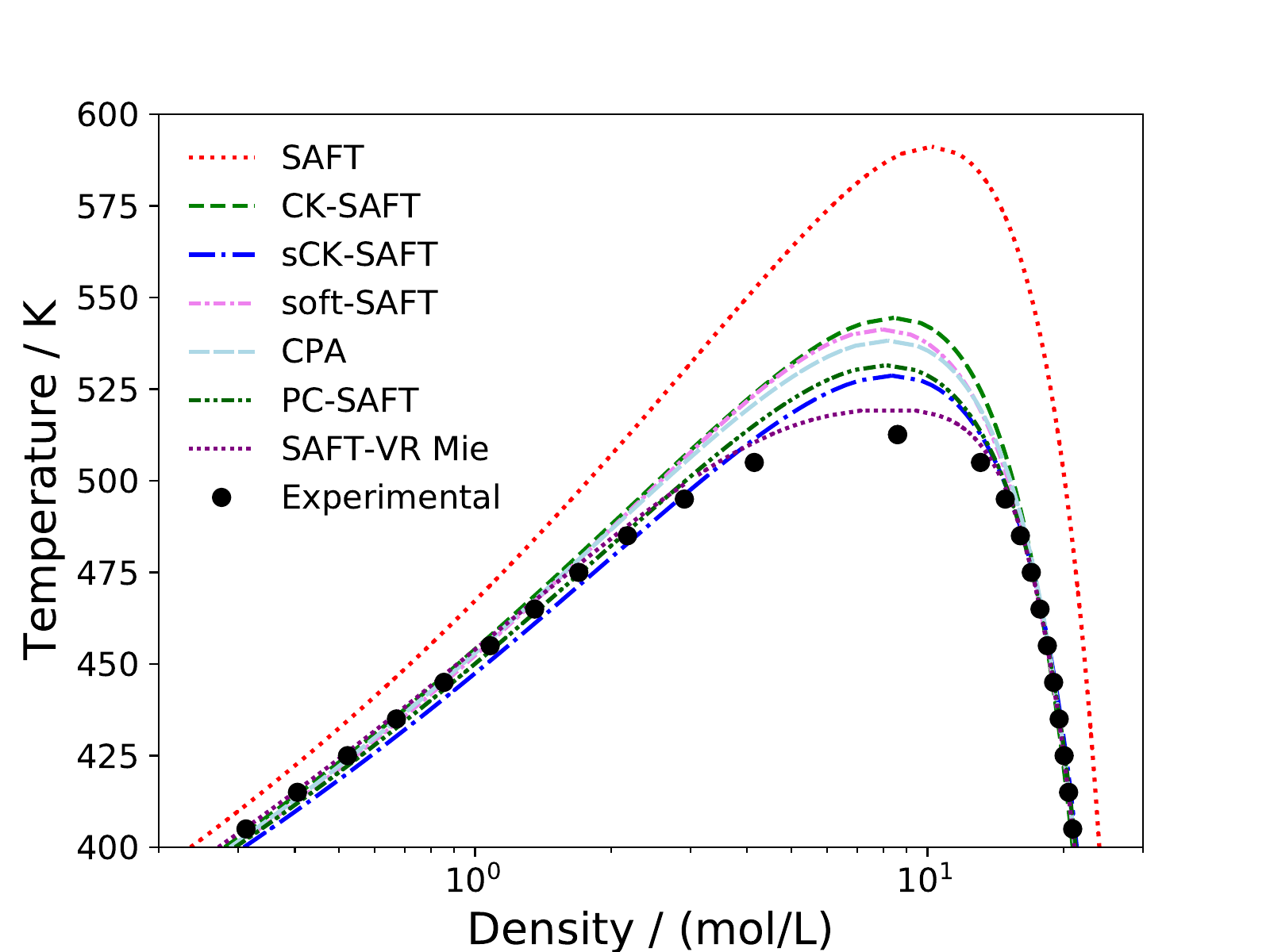}
    \caption{VLE envelope of methanol for using different SAFT equations.\citep{Lemmon2020ThermophysicalSystems}}
    \label{fig:meoh_vle_saft}
  \end{subfigure}
  \begin{subfigure}[b]{0.49\textwidth}
    \includegraphics[width=1\textwidth]{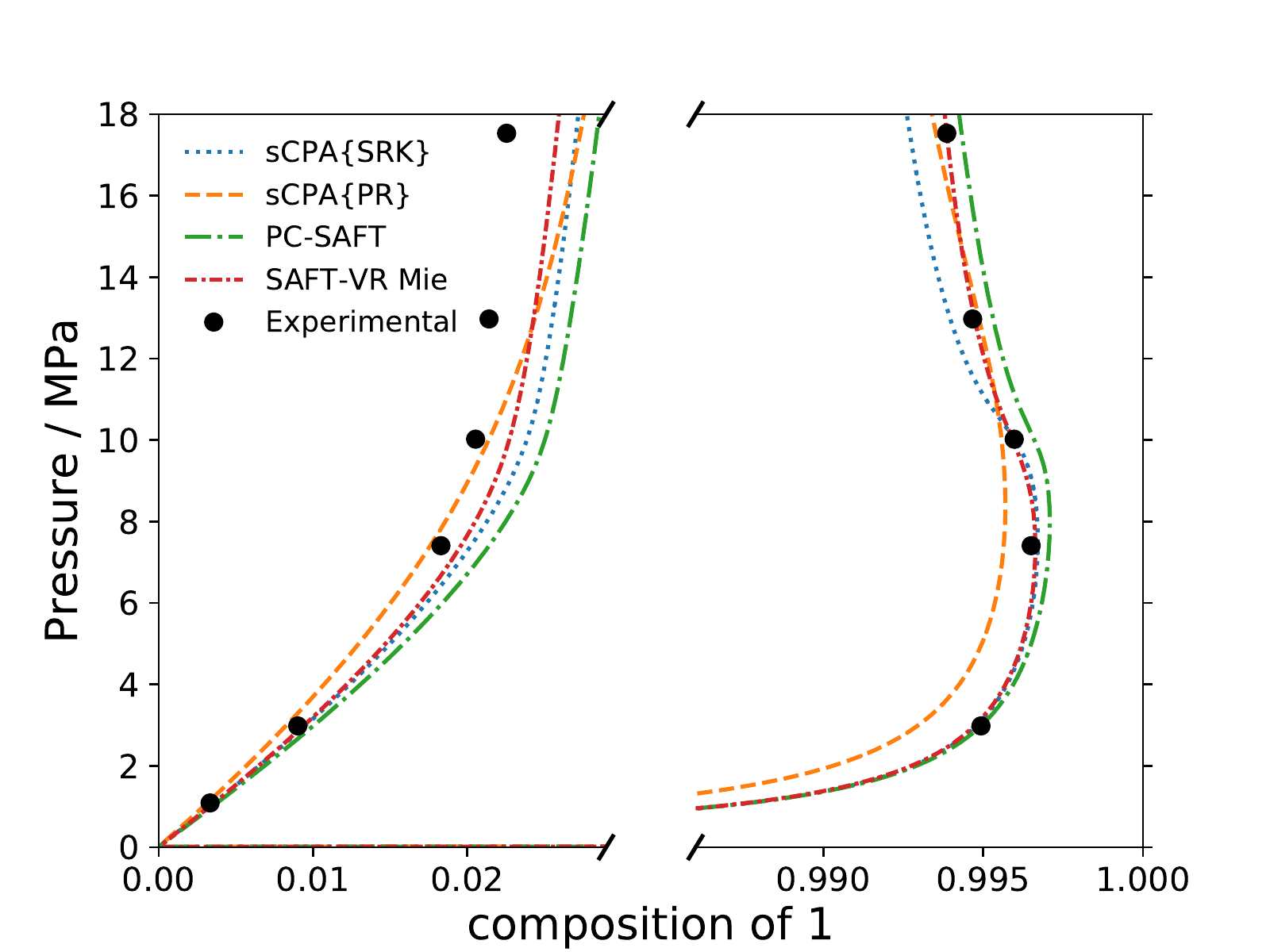}
    \caption{$pxy$ diagram of CO$_2$ (1) + water (2) at 333.15\,K using different SAFT equations.\citep{Bamberger2000High-pressureK}}
    \label{fig:water_co2_vle}
  \end{subfigure}
  \caption{Examples of the various SAFT equations available in \texttt{Clapeyron.jl}.}
\end{figure}
\noindent
\texttt{Clapeyron.jl} provides the most extensive list of SAFT-type equations of state in any open-source or commercial package available. As an example, most of the supported SAFT equations are included in figure \ref{fig:meoh_vle_saft}; methanol was selected as a species here as it requires the use of all three terms in the SAFT equation. For future reference, we remind the readers of the general formulation of the SAFT equation\citep{Chapman1990NewLiquids}:
\begin{equation}
    A_\text{res.} = A_\text{seg.}+A_\text{chain}+A_\text{assoc.}\,,
\end{equation}
where $A$ is the Helmholtz free energy and the subscripts res., seg., chain and assoc. refer to the residual contribution, the contribution due to the presence of segments, the contribution to the formation of chains of segments and the contribution due to association between segments, respectively. Whilst most SAFT equations, at the surface, appear the same, this is where most of the similarities end. The one commonality between all SAFT equations is the association term. As a result, we have common association solver used by all SAFT-type models. Here, the association fraction of sites $a$ on species $i$, $X_{ia}$, is solved for using the following mass-action equation:

\begin{equation}
    X_{ia} = \frac{1}{1 +\sum_{j,b}{n_{jb}X_{jb}\Delta_{ijab}}\rho z_{i}}\,,
\end{equation}
\noindent
where $\Delta_{ijab}$ is the association strength between site $a$ in component $i$ and site $b$ in component $j$. As this expression is identical in all SAFT equations, all of them can use one common solver. Our implementation requires only the definition of the association strength. In the general case, the mass-action equation is solved successively until convergence. However, analytical solutions do exist for special cases, for example, when there is only one pair of sites.

% (two sites or one self-associating site):

% \begin{align}
% k_{ia} &= \rho n_{ia}z_{i}\Delta_{ijab}\,, \\
% k_{jb} &= \rho n_{ib}z_{j}\Delta_{ijab}\,, \\
% 0 &= k_{ia}x^2 + (1-k_{ia}+k_{jb})x -1\,, \\
% X_{ia} &= \frac{2}{1-k_{ia}+k_{jb} + \sqrt{(1-k_{ia}+k_{jb})^2 +4k_{ia}}}\,, \\
% X_{jb} &= \frac{1}{1+k_{ia}X_{ia}} \,.
% \end{align}

Aside from the common association term, all other terms in SAFT equations are too distinct to offer the same level of flexibility present in the cubic equations. Nevertheless, we are able to provide certain levels of flexibility to some of the SAFT equations:
\begin{enumerate}
    \item Variants: This feature is not exclusive to SAFT-type equations and we have given an example in section \ref{sect:arch}. However, it has been used to provide variants of SAFT-type equations for multiple base SAFT equations. These have been given in table \ref{tbl:models}. There is one other equation of state we treat as a `variant', although not in the same sense as the example given previously.  It is possible to treat SAFT-$\gamma$ Mie as a variant of SAFT-VR Mie as they share practically the same equations\citep{Lafitte2013AccurateSegments,Papaioannou2014GroupSegments} only adapted for a group-contribution method. To take advantage of this fact, within our implementation of SAFT-$\gamma$ Mie, we map this equation to the SAFT-VR Mie equation in the following way (we refer to the mapped model as \texttt{vrmodel}):
    \begin{itemize}
        \item Segment term: Within the segment term of SAFT-$\gamma$ Mie, groups are treated as separate entities. As a result one can simply treat the groups as species when mapping from SAFT-$\gamma$ Mie to SAFT-VR Mie.
        \item Chain term: Since the groups are combined to make a single chain in SAFT-$\gamma$ Mie, the group-based parameters are averaged to obtain effective species-based parameters. Aside from this averaging, the formulation of the chain terms in both equations is identical. As such, one can simply feed the effective parameters straight into the SAFT-VR Mie chain term.
        \item Association term: This mapping is a little more difficult as variables are related to the species, groups and sites. To handle this, the groups are effectively ignored and the sites that are present are then linked to the species. This means that we must introduce new labels for the association sites, rather than just the standard `hydrogen' (\texttt{H}) or `electron' (\texttt{e1} or \texttt{e2}) sites\citep{Lafitte2013AccurateSegments,Papaioannou2014GroupSegments}; these labels now contain the groups on which the sites are present. As an example:
        \begin{minted}[breaklines,escapeinside=||,mathescape=true,  numbersep=1pt, gobble=2, frame=lines, fontsize=\small, framesep=2mm]{julia}
julia> model = SAFTgammaMie([
("ethanol",["CH3"=>3,"CH2OH"=>1]),
("glycolic acid", ["COOH"=>1,"CH2OH"=>1])]);

julia> model.sites
SiteParam with 3 components:
 "CH3": (no sites)
 "CH2OH": "H" => 1, "e1" => 2
 "COOH": "e2" => 2, "H" => 1, "e1" => 2

julia> model.vrmodel.sites
SiteParam with 2 components:
 "ethanol": "CH2OH{H}" => 1, 
            "CH2OH{e1}" => 2
 "glycolic acid": "CH2OH{H}" => 1,
                  "CH2OH{e1}" => 2, 
                  "COOH{e2}" => 2, 
                  "COOH{H}" => 1, 
                  "COOH{e1}" => 2
    \end{minted}
    We can see that three groups with three site types have been mapped to two species with five site types, making the model compatible with the SAFT-VR Mie equation.
    \end{itemize}
    This unconventional use of multiple-dispatch has allowed for a simpler implementation of what is usually accepted as one of the most-complex SAFT equations. 
    \item Cubic-plus-association (CPA): The CPA equation of state\citep{Kontogeorgis1996AnFluids} is rather unique, in contrast to other SAFT-type equations. Whilst it retains the association term, the segment and chain terms are replaced by the residual Helmholtz free energy obtained from a cubic equation of state (in its original publication, SRK is used to provide this contribution). Due to the flexibility of the cubic equations of state, in literature, this flexibility is often extended to the CPA equation. Similarly, within \texttt{Clapeyron.jl}, we have extended the flexibility available for cubics to the CPA equation of state. For example:
    \begin{minted}[breaklines,escapeinside=||,mathescape=true,  numbersep=1pt, gobble=2, frame=lines, fontsize=\small, framesep=2mm]{julia}
julia> model = CPA(["water","ethanol"])
CPA{BasicIdeal, RK{BasicIdeal, CPAAlpha, NoTranslation, vdW1fRule}} with 2 components:
 "water"
 "ethanol"
Contains parameters: a, b, c1, Tc, epsilon_assoc, bondvol, Mw

julia> model = CPA(["water","ethanol"];
    cubicmodel=PR,
    translation=RackettTranslation,
    mixing=MHV1Rule,
    activity=UNIFAC)
CPA{BasicIdeal, PR{BasicIdeal, CPAAlpha, RackettTranslation, MHV1Rule{UNIFAC}} with 2 components:
 "water"
 "ethanol"
Contains parameters: a, b, c1, Tc, epsilon_assoc, bondvol, Mw
    \end{minted}
    This has resulted in possibly the most-flexible implementation of the CPA equation of state, even including commercial packages. For example, as illustrated in figure \ref{fig:water_co2_vle}, one can switch the cubic equation that is used in simplified CPA (a variant of CPA).
\end{enumerate}
\subsection{Activity-coefficient models}
\begin{figure}[h!]
\centering
  \begin{subfigure}[b]{0.49\textwidth}
    \includegraphics[width=1\textwidth]{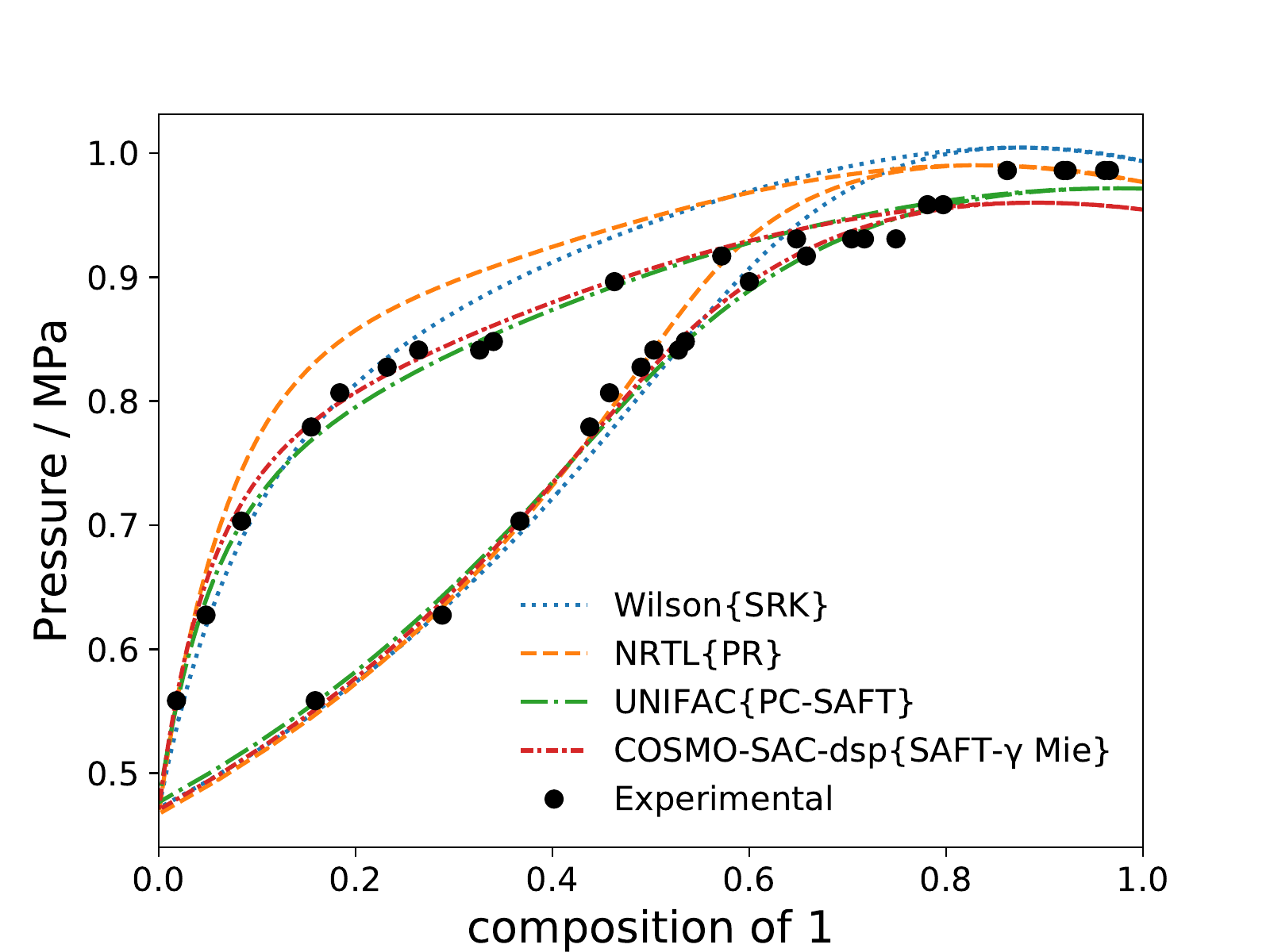}
    \caption{$pxy$ diagram of ethanol (1) + water (2) at 433.15\,K.\citep{Barr-David1959Vapor-Liquid2-Propanol-Water.}}
    \label{fig:etoh_h2o_pxy}
  \end{subfigure}
  \begin{subfigure}[b]{0.49\textwidth}
    \includegraphics[width=1\textwidth]{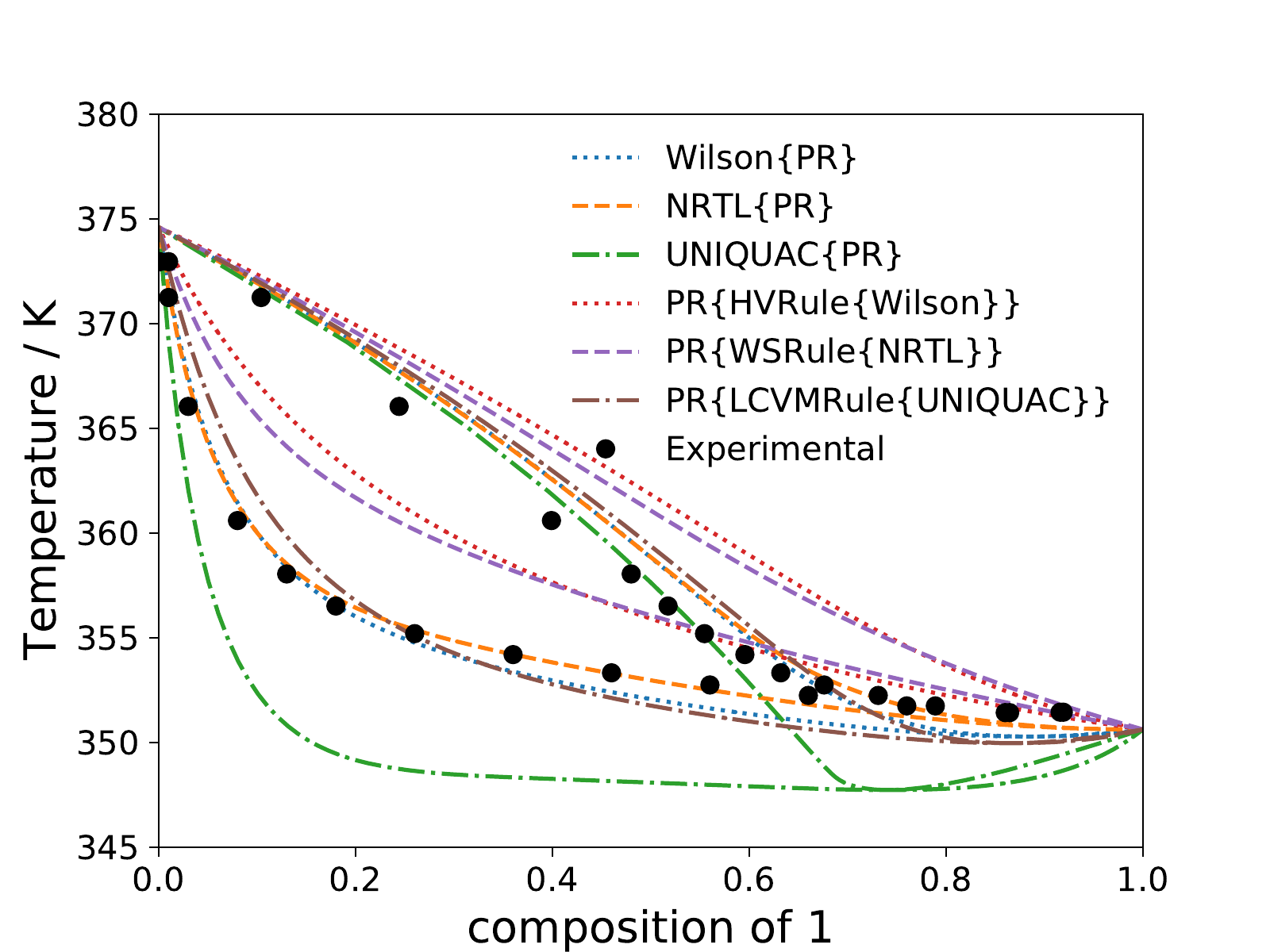}
    \caption{$Txy$ diagram of ethanol (1) + water (2) at 1.01\,bar.\citep{Stabnikov1972OnPressure}}
    \label{fig:etoh_h2o_Txy}
  \end{subfigure}
  \caption{Examples of the different capabilities of \texttt{Clapeyron.jl} for activity-coefficient models.}
\end{figure}

\noindent
Activity-coefficient models are a unique form of thermodynamic models, in contrast to the equations of state discussed previously. Whilst we have already shown them used within mixing rules to improve the predictions made by cubic equations of state in multi-component mixtures, they can be used directly to give us the activity coefficients of a mixture using which we can perform vapour--liquid equilibrium calculations through a modified Raoult's law:
\begin{equation}
    y_ip = x_i\gamma_ip_{\text{sat},i}\,,
    \label{eq:Mod_Raoult}
\end{equation}
\noindent
where $p_{\text{sat},i}$ is the saturation pressure of component $i$. Unfortunately, activity-coefficient models cannot obtain the saturation pressure of individual components themselves; this must be provided by another model. To enable this, all activity-coefficient models have an additional input to select which equation of state will be used to determine the pure-component saturation pressure:
\begin{minted}[breaklines,escapeinside=||,mathescape=true,  numbersep=1pt, gobble=2, frame=lines, fontsize=\small, framesep=2mm]{julia}
julia> model = NRTL(["methanol","benzene"];
puremodel=PR)
NRTL{PR} with 2 components:
 "methanol"
 "benzene"
Contains parameters: a, b, c, Mw
    \end{minted}
Any equation of state can be used to obtain the saturation pressure, as long as all component parameters are available; this is illustrated in figure \ref{fig:etoh_h2o_pxy}. Unfortunately, many of the limitations of activity-coefficient models remain:
\begin{itemize}
    \item Sub-critical regions only: Given activity-coefficient models rely on Raoult's law to obtain vapour--liquid equilibrium properties, it is required that all components in the system be below their critical point. Thus, activity model calculations are limited to the lowest critical temperature of the components in the mixture. 
    \item Bulk properties: activity-coefficient models are solely designed for vapour--liquid equilibria calculations and, technically, cannot be used to obtain bulk properties. However, using Dalton's law as an approximation, one could obtain an expression for the total Helmholtz free energy of the system:
    \begin{equation}
        A(V,T,\mathbf{z})= G^E(T,\mathbf{z})+G^I(T,\mathbf{z})+\sum_iz_i(G_i(V,T)-p_i(V,T)V)\,.
    \end{equation}
    where $G^E$, $G^I$, $G_i$ and $p_i$ are the excess Gibbs free energy, ideal Gibbs free energy of mixing, Gibbs free energy of pure component $i$ and pressure of pure component $i$, respectively. Although this provides a route to the provision of bulk properties, it is, of course, a very crude approximation and is not recommended.
\end{itemize}
Nevertheless, with the level of flexibility provided by \texttt{Clapeyron.jl}, one can produce interesting comparisons between various activity-coefficient models when they are used by themselves or within a mixing rule (as shown in figure \ref{fig:etoh_h2o_Txy}). 

\subsection{Ideal gas models}
Often overlooked as they have no impact on equilibrium properties, the ideal contribution to any equation of state can represent the dominant contribution to various properties, particularly second-derivative properties.\citep{Walker2020AState} As such, within \texttt{Clapeyron.jl}, we've provided both basic and more-advanced ideal contribution models. 

\begin{figure}[h!]
\centering
  \begin{subfigure}[b]{0.49\textwidth}
    \includegraphics[width=1\textwidth]{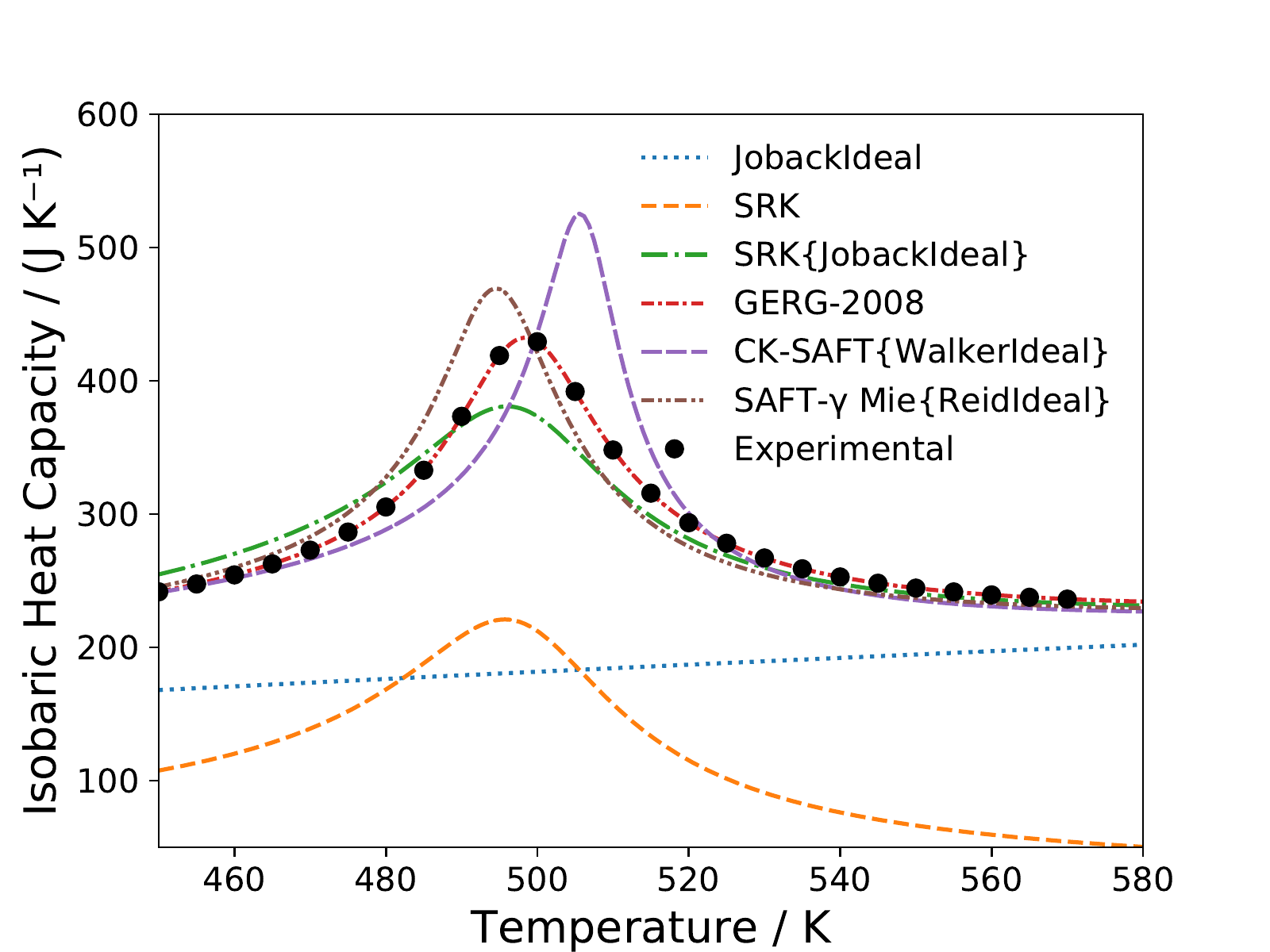}
    \caption{Isobaric heat capacity of pentane at \\ 5\,MPa.\citep{Lemmon2020ThermophysicalSystems}}
    \label{fig:pentane_cp}
  \end{subfigure}
  \begin{subfigure}[b]{0.49\textwidth}
    \includegraphics[width=1\textwidth]{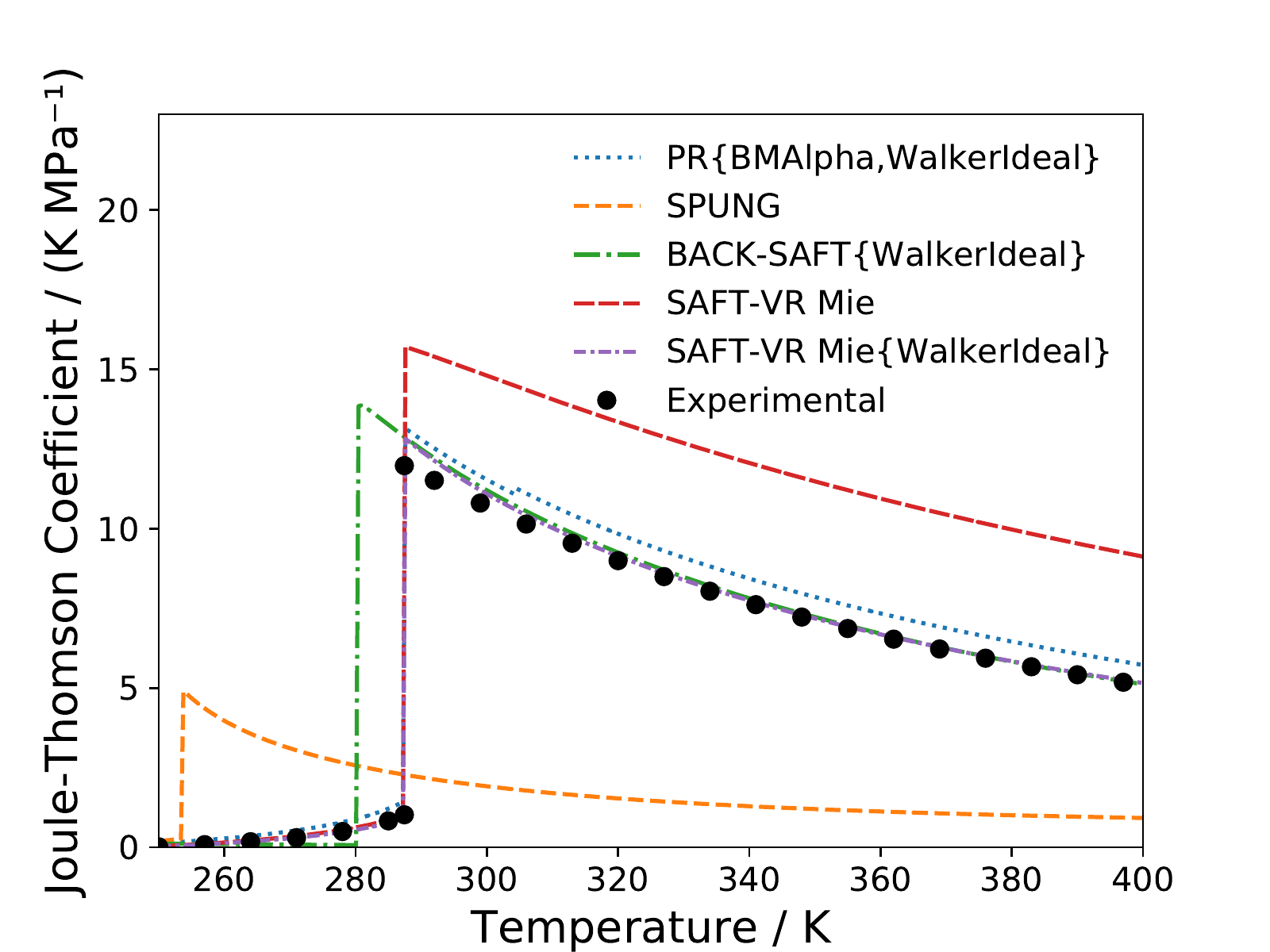}
    \caption{Joule--Thomson coefficient of CO$_2$ at 5\,MPa.\citep{Lemmon2009ThermodynamicMPa}}
    \label{fig:co2_JT}
  \end{subfigure}
  \caption{Examples of different applications of the ideal term within \texttt{Clapeyron.jl}.}
\end{figure}

Firstly, astute readers may have noticed that, whenever a model is created, one of the sub-types it contains is \texttt{BasicIdeal}. This is the default ideal model which requires no species-based parameters. It is given by the following equation:
\begin{equation}
    A_\text{ideal} = Nk_\text{B}T\sum_i x_i\left[\log{\left(\frac{x_i}{VT^{3/2}}\right)}-1\right]\,.
\end{equation}
This equation is only meant to account for the translation modes of species, thus, will give quantitatively inaccurate results for bulk properties such as the heat capacities (see the results obtained from SRK in figure \ref{fig:pentane_cp} and SAFT-VR Mie in figure \ref{fig:co2_JT}), except for molecules (such as noble gases) where only translation is important. To remedy this, we have provided three alternative ideal terms: Reid polynomials\citep{Reid1959TheLiquids}, the Joback method\citep{JOBACK1987ESTIMATIONGROUP-CONTRIBUTIONS} and Walker's model\citep{Walker2020AState}. All three of these will account for the rotational and vibrational modes missing from the \texttt{BasicIdeal} model. For example:
\begin{minted}[breaklines,escapeinside=||,mathescape=true,  numbersep=1pt, gobble=2, frame=lines, fontsize=\small, framesep=2mm]{julia}
julia> model=SAFTVRMie(["benzene"])
SAFTVRMie{BasicIdeal} with 1 component:
 "benzene"
Contains parameters: segment, sigma, lambda_a, lambda_r, epsilon, epsilon_assoc, bondvol, Mw

julia> isobaric_heat_capacity(model,1e5,298.15)
64.24460509411827

julia> model=SAFTVRMie(["benzene"];idealmodel=WalkerIdeal)
SAFTVRMie{WalkerIdeal} with 1 component:
 "benzene"
Contains parameters: segment, sigma, lambda_a, lambda_r, epsilon, epsilon_assoc, bondvol, Mw

julia> isobaric_heat_capacity(model,1e5,298.15)
125.6270421887885
    \end{minted}
Note that, in the above, we have combined a non-group-contribution residual model with a group-contribution ideal model. Such combinations are allowed in \texttt{Clapeyron.jl} as long as the groups for that particular species are available within \texttt{Clapeyron.jl}. The opposite is also possible (combining a group-contribution residual model with a non-group-contribution ideal model). Both of these situations are illustrated in figures \ref{fig:pentane_cp} and \ref{fig:co2_JT} where readers can see that, even for properties like the Joule--Thomson coefficient, the ideal model can have a significant impact on the overall property. Figure \ref{fig:pentane_cp} also highlights that one can obtain results from the ideal models alone in \texttt{Clapeyron.jl}.

Whilst the current library of ideal terms is smaller than the other models, the terms that are included provide extensive databases of parameters which should be compatible with almost any species. If more ideal terms are needed, these will be added over time or can be provided by the users themselves.

\subsection{Empirical Helmholtz equations}
What sets these equations apart from the rest is that they do not have a common structure between them and are designed with particular systems in mind. They are designed to be highly accurate for those particular systems (see figures \ref{fig:pentane_cp} and \ref{fig:Hvap_water} as examples for GERG-2008\citep{Kunz2012TheGERG-2004}). 

The lack of a common structure and the correlative nature of these equations prevents us from developing flexible frameworks in which to use them (save for the SPUNG framework). Nevertheless, they serve as an example that thermodynamic models within \texttt{Clapeyron.jl} need not belong to the cubic, SAFT or activity-coefficient family of models. They can be stand-alone and still be compatible with \texttt{Clapeyron.jl}'s solvers.
\subsubsection{SPUNG}
The State Research Program for Utilization of Natural Gas (SPUNG) framework was initially devised by \citet{Jrstad1993EquationHydrocarbons} based on the work by \citet{Mollerup1998UnificationStates} using the principle of corresponding states. The general idea here is, if we have a model that is very good at representing a specific system, according to the principle of corresponding states, we should be able to map the results of this model to another similar model and expect to achieve similarly accurate results. The details of this are explained in the original works. As one can see in figure \ref{fig:pentane_cp}, if the two species are dissimilar, such as carbon dioxide and propane, the results obtained can be quite poor.

This framework requires two things: 1. A highly accurate reference model for the reference system; 2. A `shape' model, which is used to map the reference system to the desired system. Taking the following example:
\begin{minted}[breaklines,escapeinside=||,mathescape=true,  numbersep=1pt, gobble=2, frame=lines, fontsize=\small, framesep=2mm]{julia}
julia> model = SPUNG(["propylene"])
Extended Corresponding States model
 reference model: PropaneRef()
 shape model: RK{BasicIdeal, SoaveAlpha, NoTranslation, vdW1fRule}("propylene")

julia> sat_pure(model,250)
(299496.50354575133, 7.191991655304882e-5, 0.006575782239022821)
    \end{minted}
As one can see, the default reference and shape models are the propane reference and SRK equations of state. Where we begin seeing the true flexibility of \texttt{Clapeyron.jl} is that the reference and shape equations can be any of the models available, leading to an unimaginable number of possibilities:
\begin{minted}[breaklines,escapeinside=||,mathescape=true,  numbersep=1pt, gobble=2, frame=lines, fontsize=\small, framesep=2mm]{julia}
# SPUNG(components,refmodel,shapemodel,shaperef)
julia> model = SPUNG(["methanol"],
                    IAPWS95(),
                    PCSAFT(["methanol"]),
                    PCSAFT(["water"]))
Extended Corresponding States model
 reference model: IAPWS95("water")
 shape model: PCSAFT("methanol")

 julia> model = SPUNG(["water"],vdW(["methane"]),
                    SAFTgammaMie(["water"]),
                    SAFTgammaMie(["methane"]))
Extended Corresponding States model
 reference model: vdW("methane")
 shape model: SAFTgammaMie("water")
    \end{minted}
The latter example is bound to be inaccurate but helps illustrate just how flexible \texttt{Clapeyron.jl} can be. 
\subsection{User-defined equations}
The last class of thermodynamic models supported in \texttt{Clapeyron.jl} are those defined by the users themselves. To limit the barrier of entry to novice users, we have provided generic templates within our documentation. However, this can summarised to just four components:
\begin{minted}[breaklines,escapeinside=||,mathescape=true,  numbersep=1pt, gobble=2, frame=lines, fontsize=\small, framesep=2mm]{julia}
# Defining the parameters used by the model
struct NewEoSParam <: EoSParam
    # Insert parameter names here along with their type (SingleParam, PairParam or AssocParam)
end

# Defining an abstract type for this model type
abstract type NewModel <: SAFTModel end

# Creating a model struct called New, which is a sub-type of NewEoSModel, and uses parameters defined in NewEoSParam
@newmodel NewEoS NewEoSModel NewEoSParam

# Function that will create the model object
function NewEoS(components; idealmodel=BasicIdeal, userlocations=String[], ideal_userlocations=String[], verbose=false)
    # Obtain a Dictionary of parameters. We pass in custom locations through the optional parameter userlocations.
    params = getparams(components; userlocations=userlocations, verbose=verbose)
    
    # Insert any necessary transformation to the parameters (e.g. combining rules)
    
    # Create the parameter struct that we have defined.
    packagedparams = NewEoSParam("transformed parameters go here")
    
    # Although optional, it's generally good practise to cite your models!
    references = ["DOI 1", "DOI 2"]

    model = NewEoS(packagedparams, idealmodel; ideal_userlocations=ideal_userlocations, references=references, verbose=verbose)
  
    # Return the model object that you have just created.
    return model
end

# Defining the equations
function a_res(model::NewEoSModel, V, T, z)
       # Define all necessary equations that output the residual Helmholtz free energy here
       return a_res
end
    \end{minted}

However, this is assuming the desired model is entirely separate from all those defined in \texttt{Clapeyron.jl}. If the user wishes to define a variant of an existing thermodynamic model in \texttt{Clapeyron.jl}, the only equations they would need to define are those that are different from the original equation coded in \texttt{Clapeyron.jl}. This is a similar concept to how simplified PC-SAFT was implemented in \texttt{Clapeyron.jl}, only needing to redefine the modified equations. A full example of this is given in supplementary material for the SAFT-VR Mie+AT equation developed by \citet{WalkerAbState}, requiring less than 150 lines to implement (compared to the 900 lines needed to code the SAFT-VR Mie equation).

In general, to assure compatibility with \texttt{Clapeyron.jl}, it is recommended to simply modify existing equations in \texttt{Clapeyron.jl} or, at the very least, have the new equation be a sub-type of \texttt{CubicModel}, \texttt{SAFTModel} or \texttt{ActivityModel}. This increases the chance that the existing methods in \texttt{Clapeyron.jl} will instantly be compatible with the new model. In the following section we illustrate why this is the case.
\section{Supported Methods}
\label{sect:methods}
In this section we highlight some of the supported methods in \texttt{Clapeyron.jl} which are listed in table \ref{tbl:props}. These are intended to cover a wide range of properties whilst also being intuitive to use. Many of the standard properties are provided although some unusual ones are available, such as the critical points for both pure and mixture systems. These are rarely seen in commercial packages such as gPROMS\citep{2020GPROMS} or ASPEN\citep{2016AspenPlus}. Details of their implementation will be given in subsequent sections. However, all of these properties are compatible with all models given in table \ref{tbl:models} (with obvious exceptions such activity models not being compatible with single-component VLE properties or critical properties). 

We note that the methods currently supported in \texttt{Clapeyron.jl} may not be as robust as those provided in commercial packages, and we will point out some of the short-comings of these methods, as well as some potential solutions to these short-comings. However, as \texttt{Clapeyron.jl} is an open-source initiative, we hope others in the field will be generous enough to contribute to the package, continuously improving its methods.

All units are stated in MKS units by default. However, \texttt{Clapeyron.jl} is also fully compatible with the \texttt{Unitful.jl} package\citep{Keller2022Unitful.jl}. We are effectively able to inject units to the inputs, which \texttt{Clapeyron.jl} will automatically convert into MKS units, and obtain the outputs with the correct units; this may be useful when this package is used as part of a larger system with \texttt{Unitful.jl} support. For example:
\begin{minted}[breaklines,escapeinside=||,mathescape=true,  numbersep=1pt, gobble=2, frame=lines, fontsize=\small, framesep=2mm]{julia}
julia> using Unitful

julia> import Unitful: bar, °C, mol

julia> model = PCSAFT(["methanol","ethanol"])
PCSAFT{BasicIdeal} with 2 components:
 "methanol"
 "ethanol"
Contains parameters: Mw, segment, sigma, epsilon, epsilon_assoc, bondvol

julia> Cp = isobaric_heat_capacity(model, 5bar, 25°C, [0.5mol, 0.5mol];output=cal*K^-1)
15.091222228910153 cal K⁻¹
\end{minted}

\begin{table*}[h!]
\caption{List of supported properties within \texttt{Clapeyron.jl}, along with their function names.}
\label{tbl:props}
\resizebox{\textwidth}{!}{
\begin{tabular}{lll}
\textbf{Type}                                                 & \textbf{Property}                   & \textbf{Function}                                   \\ \hline
\multirow{19}{*}{Bulk}                               & Volume                     & \texttt{V=volume(model, p, T, z)}                   \\ \cline{2-3} 
                                                     & Entropy                    & \texttt{S=entropy(model, p, T, z)}                  \\ \cline{2-3} 
                                                     & Chemical Potential         & μ\texttt{=chemical\_potential(model, p, T, z)}         \\ \cline{2-3} 
                                                     & Internal Energy            & \texttt{U=internal\_energy(model, p, T, z)}            \\ \cline{2-3} 
                                                     & Enthalpy                   & \texttt{H=enthalpy(model, p, T, z)}                   \\ \cline{2-3} 
                                                     & Gibbs Free Energy          & \texttt{G=gibbs\_free\_energy(model, p, T, z)}          \\ \cline{2-3} 
                                                     & Helmholtz Free Energy      & \texttt{A=helmholtz\_free\_energy(model, p, T, z)}      \\ \cline{2-3} 
                                                     & Isochoric Heat Capacity    & \texttt{Cv=isochoric\_heat\_capacity(model, p, T, z)}    \\ \cline{2-3} 
                                                     & Isobaric Heat Capacity     & \texttt{Cp=isobaric\_heat\_capacity(model, p, T, z)}     \\ \cline{2-3} 
                                                     & Isothermal Compressibility & β\texttt{T=isothermal\_compressibility(model, p, T, z)} \\ \cline{2-3} 
                                                     & Isentropic Compressibility & β\texttt{S=isentropic\_compressibility(model, p, T, z)} \\ \cline{2-3} 
                                                     & Speed of Sound             & \texttt{u=speed\_of\_sound(model, p, T, z)}             \\ \cline{2-3} 
                                                     & Isobaric Expansivity       & α\texttt{V=isobaric\_expansivity(model, p, T, z)}       \\ \cline{2-3} 
                                                     & Joule--Thomson Coefficient & μ\texttt{JT=joule\_thomson\_coefficient(model, p, T, z)}  \\ \cline{2-3} 
                                                     & Fugacity Coefficient       & φ\texttt{=fugacity\_coefficient(model, p, T, z)}       \\ \cline{2-3} 
                                                     & Activity Coefficient       & γ\texttt{=activity\_coefficient(model, p, T, z)}       \\ \cline{2-3} 
                                                     & Compressibility Factor     & \texttt{Z=compressibility\_factor(model, p, T, z)}     \\ \cline{2-3} 
                                                     & Inversion Temperature      & \texttt{T=inversion\_temperature(model, p, z)}         \\ \cline{2-3} 
                                                     & Second virial coefficient  & \texttt{B=second\_virial\_coefficient(model, T, z)}     \\ \hline
Mixing & Mixing Function & \texttt{prop\_mix=mixing(model,p,T,z,property)}\\\hline
Excess & Excess Function & \texttt{prop\_E=excess(model,p,T,z,property)}\\\hline
\multirow{3}{*}{VLE (single-component)}            & Saturation pressure        & \texttt{p\_sat, V\_l, V\_v=saturation\_pressure(model, T)}           \\ \cline{2-3} 
                                                     & Saturation Temperature     & \texttt{T\_sat, V\_l, V\_v=saturation\_temperature(model, p)}        \\ \cline{2-3} 
                                                     & Enthalpy of Vapourisation  & \texttt{H\_vap = enthalpy\_vap(model, T)}                     \\ \hline
Critical point   (single-component)                  & Critical point             & \texttt{Tc, pc, Vc=crit\_pure(model) }                          \\ \hline
\multirow{10}{*}{VLE and VLLE   (multi-component)}   & Bubble pressure            & \texttt{p\_bub, V\_l, V\_v, y=bubble\_pressure(model, T, x)  }             \\ \cline{2-3} 
                                                     & Bubble temperature         & \texttt{T\_bub, V\_l, V\_v, y=bubble\_temperature(model, p, x) }         \\ \cline{2-3} 
                                                     & Dew pressure               & \texttt{p\_dew, V\_l, V\_v, x=dew\_pressure(model, T, y)}                \\ \cline{2-3} 
                                                     & Dew temperature            & \texttt{T\_dew, V\_l, V\_v, x=dew\_temperature(model, T, y)}             \\ \cline{2-3} 
                                                     & LLE pressure               & \texttt{p\_LLE, V\_l, V\_ll, xx=LLE\_pressure(model, T, x) }                 \\ \cline{2-3} 
                                                     & LLE temperature            & \texttt{T\_LLE, V\_l, V\_ll, xx=LLE\_temperature(model, p, x)}        \\ \cline{2-3} 
                                                     & Azeotrope pressure         & \texttt{p\_az, V\_l, V\_v, x=azeotrope\_pressure(model, T)}            \\ \cline{2-3} 
                                                     & Azeotrope temperature      &  \texttt{T\_az, V\_l, V\_v, x=azeotrope\_temperature(model, p)}         \\ \cline{2-3} 
                                                     & VLLE pressure              &  \texttt{p\_VLLE, V\_l, V\_ll, V\_v, x, xx, y=VLLE\_pressure(model, T)}                    \\ \cline{2-3} 
                                                     & VLLE temperature              & \texttt{T\_VLLE, V\_l, V\_ll, V\_v, x, xx, y=VLLE\_temperature(model, p)}                 \\ \hline
\multirow{3}{*}{Critical points   (multi-component)} & Critical point (VLE)       & \texttt{Tc, pc, Vc=crit\_mix(model, x)}                         \\ \cline{2-3} 
                                                     & UCST                       & \texttt{p\_UCST, V\_UCST, x\_UCST=UCST\_mix(model, T) }                            \\ \cline{2-3} 
                                                     & UCEP                       & \texttt{T\_UCEP, p\_UCEP, V\_l, V\_v, x, y=UCEP\_mix(model)}                                \\ \hline
\end{tabular}}
\end{table*}
\subsection{Bulk properties}
\label{sect:bulk}
Bulk properties are properties attributed to a single phase at a given pressure, temperature and composition (as summarised in table \ref{tbl:props}). We make use of the automatic differentiation (AD) provided by the \texttt{ForwardDiff.jl} package\citep{Revels2016Forward-ModeJulia} to provide the derivatives needed to obtain these properties, which ensures that all properties are compatible with all thermodynamic models. 
The package works by providing a Dual number type, that propagates the evaluation of a set of primal functions ($+$, $-$, sin, exp) with it's derivative. The derivative of an arbitrary function composed of differentiable primitives is just the propagation of those derivatives via the chain rule. This paradigm is known as differentiable programming.\citep{Ma2018ASolutions}
%cite this: arXiv:1812.01892
Because the differentiation is done at the primitive level, the error associated with the derivative is similar to the error of the original function. Furthermore, one differentiation evaluation results in the original function and its derivative, allowing the use of both values with no additional evaluation time. 

An example of how automatic differentiation is used to obtain the chemical potential is shown below:
\begin{minted}[breaklines,escapeinside=||,mathescape=true,  numbersep=1pt, gobble=2, frame=lines, fontsize=\small, framesep=2mm]{julia}
function VT_chemical_potential(model::EoSModel, V, T, z=SA[1.])
    fun(x) = eos(model,V,T,x)
    return ForwardDiff.gradient(fun,z)
end
    \end{minted}

It is important to note that these thermodynamic properties are still defined in terms of $(V, T, \mathbf{z})$. Thus, if we want to obtain properties in at a specified $(p, T, \mathbf{z})$, a solver needs to be used to make that transformation. \texttt{Clapeyron.jl} provides two methods to solve for the volume of a system:
\begin{itemize}
    \item Cubic-only method: For a cubic equation of state:
    \begin{equation}
        c_0+c_1V+c_2V^2+c_3V^3=0\,,\label{eq:cubic_vol}
    \end{equation}
    it is very easy to obtain all roots analytically. When there are three real roots, one may determine which phase is the most stable by simply checking which of the three roots corresponds to the smallest Gibbs free energy.
    \item General method: Most other equations available in \texttt{Clapeyron.jl} cannot be solved as easily as \ref{eq:cubic_vol}. In the general case, we treat this as a non-linear equation with multiple roots\citep{Privat2010AreChemicals}:
    \begin{equation}
        p_0 - p(V,T_0,\mathbf{z}_0)=0\,.
    \end{equation}
    %By default, \texttt{Clapeyron.jl} uses use a Newton-Raphson algorithm (provided by \texttt{NLSolvers.jl}) to obtain the solution to the above equation.
    
    We make use of an iterative root-finding scheme to solve for the volume. We use a modified Newton-Raphson step, effectively solving for the logarithm of the volume, primarily to improve the stability in the liquid phase as the derivatives can be very large.:
    
    \begin{align}
     \ln(V_{i+1}) = \ln(V_{i})-\Delta\,,
    \\ \Delta = (p_{0}-p(V_{i}))\beta_T(V_i)\,,
    \end{align}
    
    where $\Delta$ is the step size and $\beta_T$ is the isothermal compressibility. The derivative for the step size of the algorithm above is obtained from automatic differentiation to improve the numerical stability of this method.
    
    However, the above only describes how one obtains a single solution, without checking if it is the most stable solution. This is not a problem if we already know the phase of our system at the given conditions; this can be specified as an additional input to all bulk properties listed in table \ref{tbl:props}:
    \begin{minted}[breaklines,escapeinside=||,mathescape=true,  numbersep=1pt, gobble=2, frame=lines, fontsize=\small, framesep=2mm]{julia}
julia> model = PCSAFT(["water"]);

julia> u = speed_of_sound(model,1e5,373.15;phase=:liquid)
2142.9834287736817

julia> u = speed_of_sound(model,1e5,373.15;phase=:vapour)
510.1968680367978
    \end{minted}
    
    In order to ensure that we solve for the specified phase, an initial guess close to the `true' solution is used:
    \begin{itemize}
        \item Vapour phase: Given that the second virial coefficient can be solved for directly in \texttt{Clapeyron.jl}, it is possible to obtain a second-order Virial expansion approximation of all equations of state in \texttt{Clapeyron.jl}:
        \begin{equation}
            Z = 1+B(T)\rho\,,
        \end{equation}
        where $B(T)$ is the second virial coefficient of the system. It is very easy to obtain the roots to the above expression; this will provide us with an initial guess for the volume that is very close to the true solution, as long as the system is not too close to the critical point ($\sim0.97T_c$).
        \item Liquid phase: Developing a generalised approach to obtaining good initial guesses for the liquid volume is very difficult. However, what is generally true for all equations of state supported in \texttt{Clapeyron.jl} is that they all have a lower bound for their volume, below which the equation is undefined. A simple example in SAFT-type equations is that they (should) all be undefined above a packing fraction of one:
        \begin{equation}
            \eta = 1 = \frac{\pi}{6}\frac{N}{V}\sum_ix_im_i\sigma_i^3\,;
        \end{equation}
        it is possible to re-arrange the above expression to obtain the lower-bound for $V$ ($V_\text{lb}$). We can generally assume the liquid phase, due to its incompressibility, will have a volume to within some factor of the lower bound (i.e., $V_\text{liq.}\approx CV_\text{lb}$). Using multiple dispatch, we can provide tuned $C$ parameters for various equations of state, to give us more-reliable initial guesses for the liquid phase for various equations of state. For example:
    \end{itemize}
    \begin{minted}[breaklines,escapeinside=||,mathescape=true,  numbersep=1pt, gobble=2, frame=lines, fontsize=\small, framesep=2mm]{julia}
function x0_volume_liquid(model::SAFTVRMieModel,T,z)
    v_lb = lb_volume(model,z)
    return v_lb*1.5
end
    \end{minted}

In the case where we do not know which phase is more stable, \texttt{Clapeyron.jl} will solve for the volume twice using both initial guesses. Whichever root returns the smaller Gibbs free energy is taken to be the true root. To speed this process up, by default, the two roots are solved for in parallel. It is possible to switch this feature off using an optional argument (\texttt{threaded}) to all the bulk properties. 

However, this method is not perfect. Naturally, one of its weaknesses is that we assume we provide good initial guesses for the both phases. Whilst we have found our guesses to be generally reliable for most conditions, for more-exotic systems (mixtures with a lot of association or extremely size-asymmetric mixtures), this may not be the case. 

Furthermore, some of these equations of state will observe three seemingly stable roots (GERG-2008\citep{Kunz2012TheGERG-2004} and PC-SAFT, as pointed out by \citet{Privat2010AreChemicals}, can observe an additional root). Given \texttt{Clapeyron.jl} only solves for two roots, there is no guarantee that the third root will not be the `true' root. 

Unfortunately, these problems are difficult to address on a global scale. However, there is no guarantee either that commercial tools will not experience similar issues. The difference with an open-source package such as \texttt{Clapeyron.jl} is that, should these issues arise, they can be dealt with by the users themselves. For \texttt{Clapeyron.jl} specifically, this can be done in a simple fashion by modifying built-in functions at the front-end. For example, if the initial guess for the SAFT-VR Mie equation of state is too small, one can modify the function:
\begin{minted}[breaklines,escapeinside=||,mathescape=true,  numbersep=1pt, gobble=2, frame=lines, fontsize=\small, framesep=2mm]{julia}
import Clapeyron: x0_volume_liquid, SAFTVRMieModel, lb_volume
function x0_volume_liquid(model::SAFTVRMieModel,T,z)
    v_lb = lb_volume(model,z)
    return v_lb*2.0
end
    \end{minted}
    Julia will automatically over-ride the \texttt{x0\_volume\_liquid} function, instead using the one provided by the user. This is an important advantage of open-source packages; as the package is used more often, more issues are raised and addressed, quickly. We hope, over time, more users will contribute to the package, gradually improving the robustness of its solvers.
\end{itemize}

Once we have solved for the volume, all other bulk properties can be obtained using \texttt{ForwardDiff.jl}. This has already been illustrated in figures \ref{fig:pentane_cp} and \ref{fig:co2_JT} where, in the case of the latter, we can see the solver automatically detects the phase change. 

\begin{figure*}[h!]
\centering
  \begin{subfigure}[b]{0.49\textwidth}
    \includegraphics[width=1\textwidth]{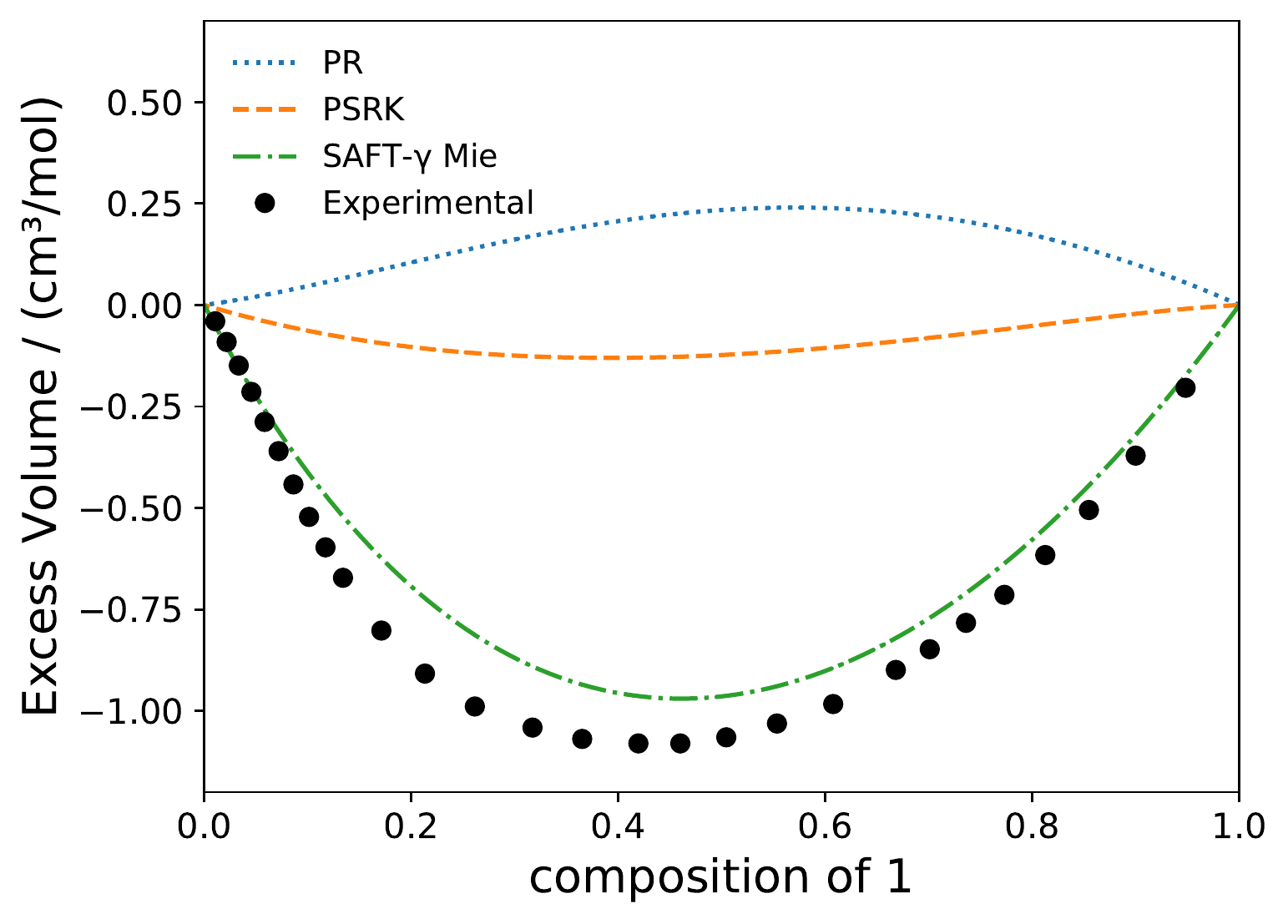}
    \caption{Excess volume of ethanol (1) + water (2) at 298.15\,K and 1.013\,bar.\citep{Ott1993ExcessMPa}}
    \label{fig:etoh_water_e_vol}
  \end{subfigure}
  \begin{subfigure}[b]{0.49\textwidth}
    \includegraphics[width=1\textwidth]{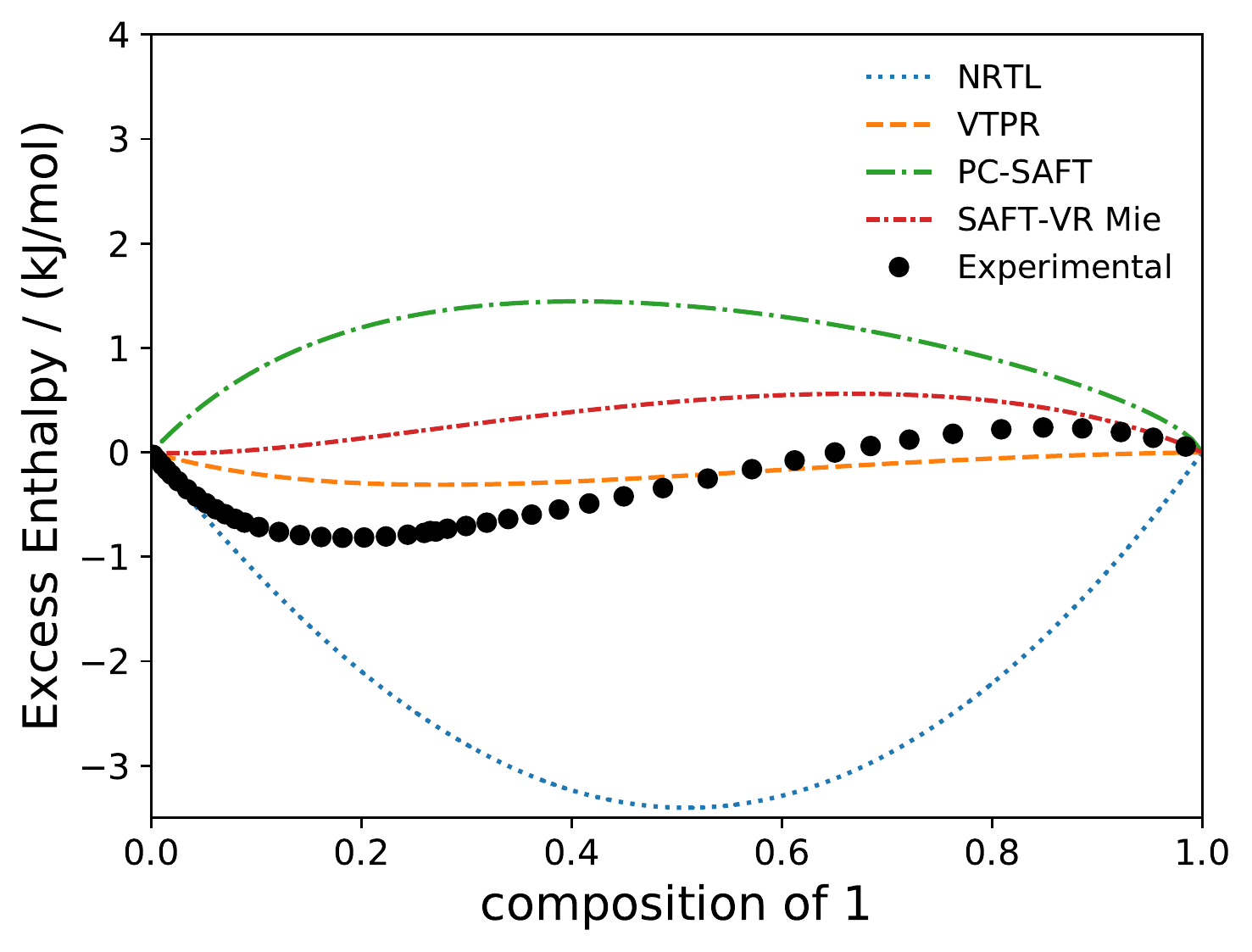}
    \caption{Excess enthalpy of acetone (1) + water (2) at 278.15\,K and 1.013\,bar.\citep{French1989ExcessK}}
    \label{fig:act_water_e_H}
  \end{subfigure}
  \caption{Mixing functions of various properties available within \texttt{Clapeyron.jl}.}
\end{figure*}

However, another type of bulk property we might be interested in is mixing and excess functions. These can be of use in continuum models such as the Cahn-Hilliard equation\citep{Inguva2021Continuum-scaleThermodynamics}. As presented in table \ref{tbl:props}, the mixing and excess functions are provided using a single function, rather than for each bulk property. This allows for users to obtain mixing or excess functions for any bulk property directly, for example:
\begin{minted}[breaklines,escapeinside=||,mathescape=true,  numbersep=1pt, gobble=2, frame=lines, fontsize=\small, framesep=2mm]{julia}
julia> model = PCSAFT(["methanol","cyclohexane"]);

julia> mixing(model, 1e5, 313.15, [0.5,0.5], volume)
1.004651584989827e-6
    \end{minted}
These functions are exemplified in figures \ref{fig:etoh_water_e_vol} and \ref{fig:act_water_e_H}.

\subsection{Pure-component VLE and critical properties}

For systems with a single species, in term of vapour--liquid equilibria properties, there are really only two types of points we are interested in finding: 

\begin{itemize}
    \item Saturation points: Here, one must find the volumes corresponding the vapour and liquid phases in equilibrium with one another which satisfy the following equations:
    \begin{align}
         p(V_l,T) &= p(V_v,T)\,,\label{eq:vle_pure1}\\
        \mu(V_l,T) &= \mu(V_v,T)\,,\label{eq:vle_pure2}
    \end{align}

    where $\mu$ is the chemical potential of a single component and the subscript $l$ and $v$ denote the liquid and vapour phases, respectively. For all equations of state of interest, the above equations represent a non-linear system of equations which needs to be solved using iterative methods\citep{Michelsen2007ThermodynamicAspects}. As such, initial guesses must be provided. We have developed two strategies to obtain these initial guesses:
    \begin{itemize}
        \item van der Waals approximation: Based on the approximations provided by \citet{Berberan-Santos2008TheSolutions}, it is possible to obtain the vapour and liquid roots for the van der Waals equation explicitly. If we map our equations of state to the van der Waals equation (a mapping that requires the second virial coefficient and the liquid volume at the specified temperature), it is then possible to obtain good initial guesses for the liquid and vapour volumes.
        \item `Vapour-like' and `liquid-like' volumes: The van der Waals approximation is not perfect and, as such, we use the approach developed in the previous section for the volume solver to obtain initial guesses for the liquid phase and vapour phase. In the case of the vapour phase, we must supply a value for the pressure which is usually chosen to be $p=-RT/(4B(T))$. This then corresponds to a vapour volume of:
        \begin{equation}
            V = -2B(T)\,.
        \end{equation}
        This approach is usually used when one is close to the critical point.
    \end{itemize}
    The combined approaches above provide a very robust saturation-pressure solver. However, if our initial guesses fail, we allow users to provide their own initial guesses using an optional argument (\texttt{saturation\_pressure(model,T,v0)}).

    We do note that, another advantage of the \texttt{ForwardDiff.jl} package is our ability to obtain the Jacobian of equations \ref{eq:vle_pure1} and \ref{eq:vle_pure2} meaning that derivative-based methods for solving linear of systems can be used, allowing for improved stability of the solvers (provided by the \texttt{NLSolvers.jl} package).

    However, the above only allows us to solve for the saturation pressure at a given temperature. In order to solve for the saturation temperature at a given pressure ($p_0$), an additional layer of iterations needs to be added. This involves solving the following equation for the temperature:
    \begin{equation}
        p_0 - p_\text{sat}(T)=0\,,
        \label{eq:p_sat_T}
    \end{equation}
    where $p_\text{sat}(T)$ is the saturation pressure at a given temperature. To solve the above equation, we use a bracketed method provided by the \texttt{Roots.jl} package; the solution is bounded between 30\% and 100\% of the critical temperature so as to ensure there is always a solution for the saturation pressure (below 30\% of the critical temperature usually corresponds to the solid phase for most species).
    \item Critical points: The critical point for a single component is defined as:
    \begin{align}
        \left(\frac{\partial^2p}{\partial V^2}\right)_T &= 0\,,\\
        \left(\frac{\partial^3p}{\partial V^3}\right)_T &= 0\,.
    \end{align}
    For cubic and empirical equations of state, the critical point of a species is usually an input parameter thus making it unnecessary to solve the above system of equations. However, for SAFT-type equations, this is not the case and we must solve the above system of non-linear equations. The third-order derivative required is seldom obtained analytically (let alone the higher order derivative required to obtain the Jacobian of the system of equations). Thus, historically, the critical point is either solved for by iteratively solving for the saturation point until $V_l=V_g$ or not supported at all (such as in gPROMS\citep{2020GPROMS}). In the case of \texttt{Clapeyron.jl}, these derivatives can all be obtained by multiple passes of forward differentiation. This leads to a visually satisfying implementation of the objective function for the critical point:
    \begin{minted}[breaklines,escapeinside=||,mathescape=true,  numbersep=1pt, gobble=2, frame=lines, fontsize=\small, framesep=2mm]{julia}
function obj_crit(model::EoSModel, F, T_c, V_c)
    _1 = one(T_c+V_c)
    ∂²A∂V², ∂³A∂V³ = ∂²³f(model, V_c, T_c, SA[_1])
    F[1] = ∂²A∂V²
    F[2] = ∂³A∂V³
    return F
end
    \end{minted}
    As we are solving for the critical volume and temperature in the above equations, we must provide initial guesses. This is done on a case-by-case basis where the critical volume is defined with respect to the lower-bound volume and the critical temperature is defined with respect to the temperature scaling (for SAFT-type equations, this is the self-interaction energy, $\epsilon/k_\text{B}$). As an example, for the general SAFT equation, the initial guess for the critical point is $T_{c,0}=2\epsilon/k_\text{B}$ and $V_{c,0}=V_\text{lb}/0.3$. Quite surprisingly, the methods used to solve the above system of equations have been found to be quite robust, regardless of chain length and presence of association. Nevertheless, we do allow for custom initial guesses to be used in case the method does fail.
    \end{itemize}
\begin{figure*}[h!]
\centering
  \begin{subfigure}[b]{0.49\textwidth}
    \includegraphics[width=1\textwidth]{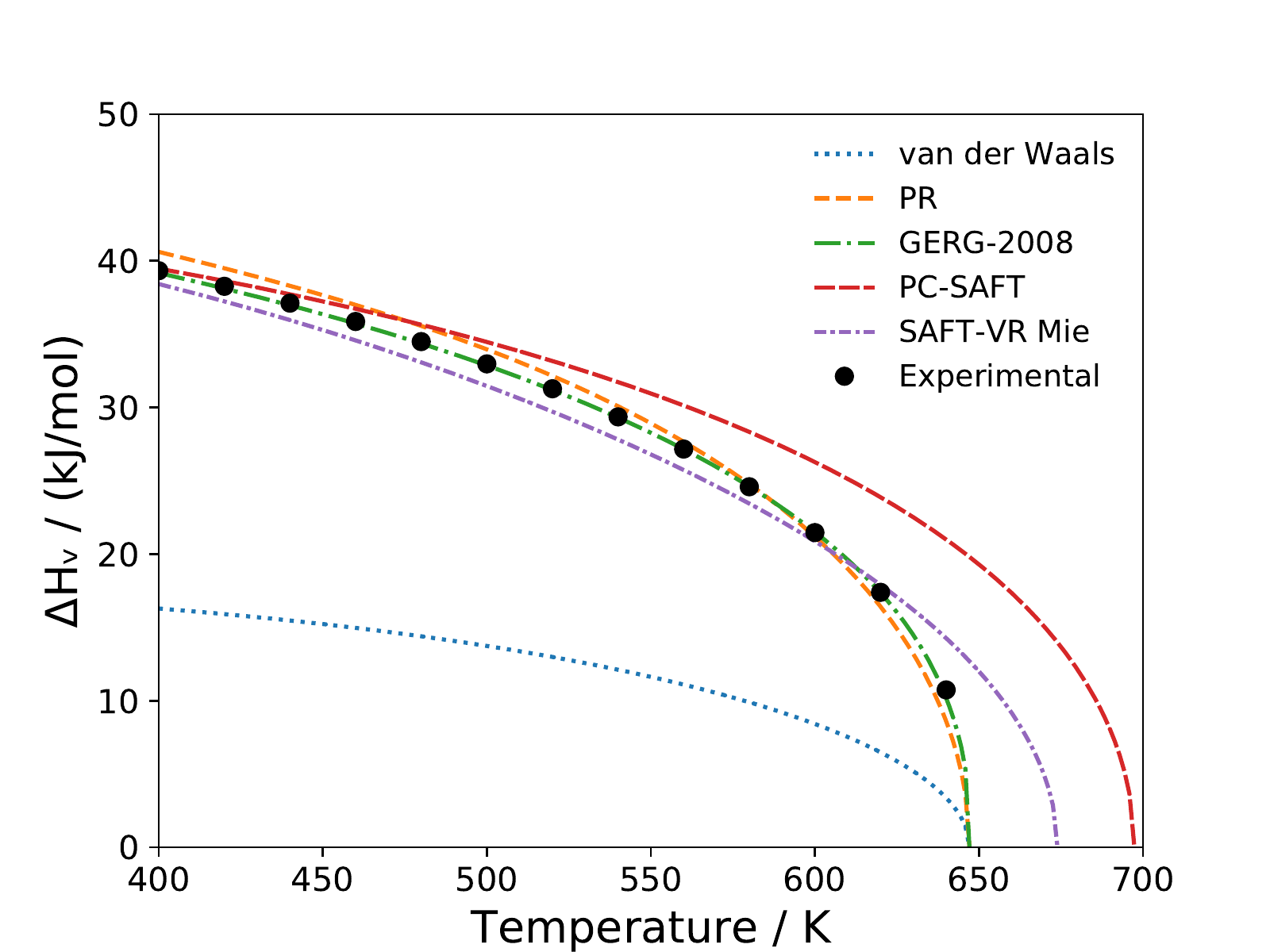}
    \caption{Enthalpy of vapourisation of water.\citep{Lemmon2020ThermophysicalSystems}}
    \label{fig:Hvap_water}
  \end{subfigure}
  \begin{subfigure}[b]{0.49\textwidth}
    \includegraphics[width=1\textwidth]{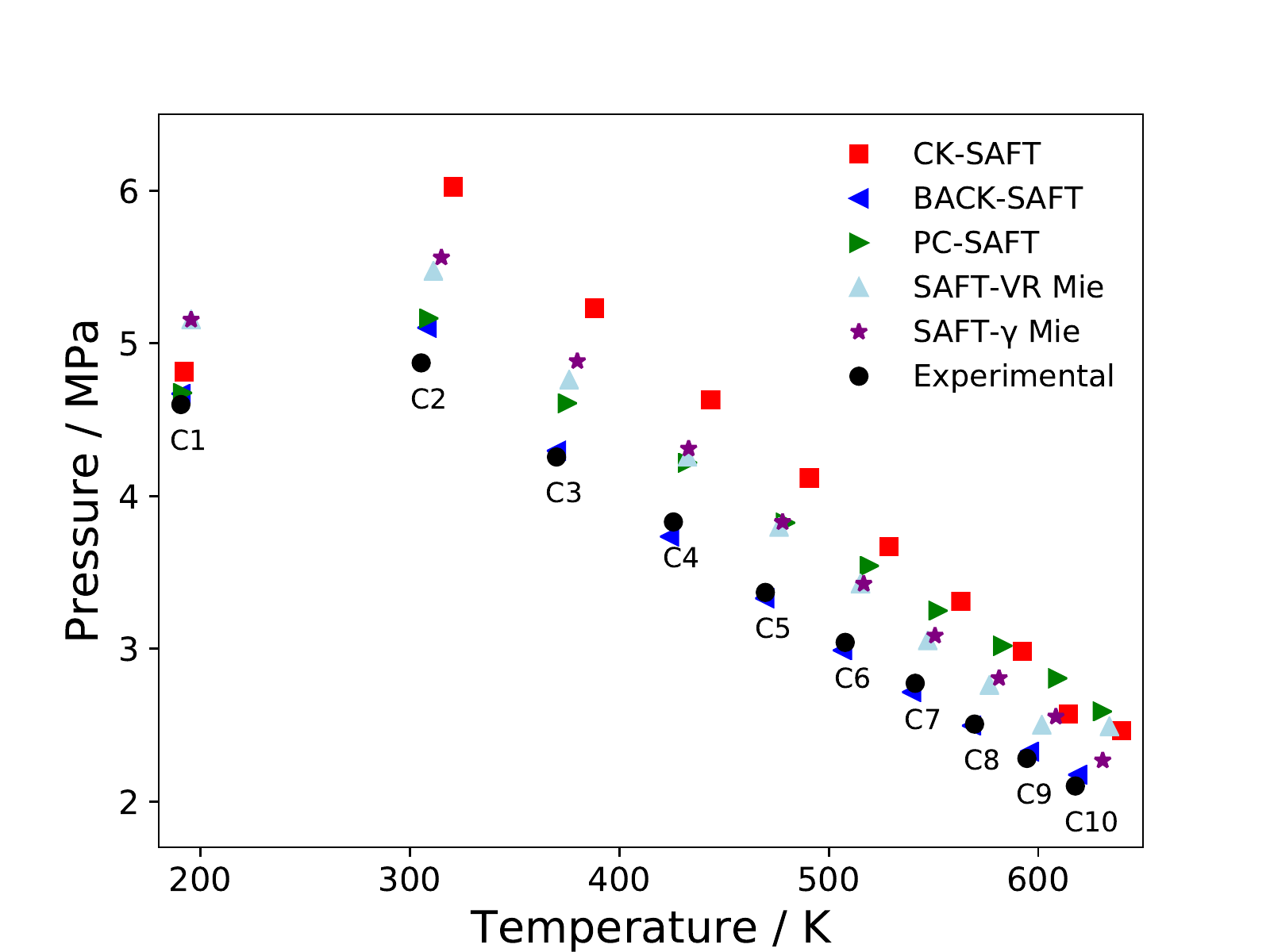}
    \caption{Critical points of various alkanes.\citep{Lemmon2020ThermophysicalSystems}}
    \label{fig:crit_alkanes}
  \end{subfigure}
  \caption{Examples of VLE and critical properties for single-component systems within \texttt{Clapeyron.jl}.}
  \label{fig:pure_VLE}
\end{figure*}
These two solvers in combination allow us to obtain the VLE envelopes of any species using any equation of state within \texttt{Clapeyron.jl}; this has already been exemplified in figures \ref{fig:water_psat_cubic}, \ref{fig:water_vle_cubic} and \ref{fig:meoh_vle_saft}. More examples of this are shown in figure \ref{fig:pure_VLE}. In figure \ref{fig:Hvap_water}, the saturation and critical point solvers are used to obtain the enthalpy of vapourisation of water from 400\,K to the critical point predicted using various equations of state. In figure \ref{fig:crit_alkanes}, the critical points for the first 10 alkanes in the homologous series are obtained using various SAFT-type equations of state; as is typically expected from these equations of state, they overestimate the critical point, with the exception of BACK-SAFT.
\subsection{Multi-component VLE and critical properties}
The last set of properties supported are the multi-component VLE, liquid--liquid equilibrium (LLE), vapour--liquid--liquid equilibrium (VLLE) and critical properties. As shown in table \ref{tbl:props}, a diverse range of these are supported in \texttt{Clapeyron.jl}. Generally, these can be divided into three types of functions:

\begin{itemize}
    \item Phase--equilibria points: These are the bubble, LLE, azeotrope and VLLE functions where two or three phase are in equilibrium with one another, satisfying the following equations:
    \begin{align}
        T^\alpha&=T^\beta&\hspace{-3.8cm}(=T^\gamma) \quad&\\
        p^\alpha&=p^\beta&\hspace{-3.8cm}(=p^\gamma) \quad&\\
        \mu_i^\alpha&=\mu_i^\beta&\hspace{-3.8cm}(=\mu_i^\gamma) \quad& \forall i\in 1,\cdots,n_c
    \end{align}

    where the superscripts $\alpha$, $\beta$ and $\gamma$ denote different phases. It is very straightforward to set-up the objective function for such a problem (regardless of which phases or how many are in equilibrium). Because of automatic differentiation, it is equally as simple to obtain all the high-order derivatives for these functions which may be needed when obtaining the roots of the problem using derivative-based methods. 
    
    The only real difficulty for these functions is providing sufficiently accurate initial guesses. However, if we assume that our saturation and critical point solvers for single component systems are robust, it is also possible to obtain reliable initial guesses:
    \begin{itemize}
        \item Bubble and dew points: In this situation, to obtain the initial guess, we make the assumption that the system can be approximated as an ideal mixture, obeying Raoult's law:
        \begin{equation}
            py_i = p_{\text{sat},i}(T)x_i\,,
        \end{equation}
        we can easily obtain $p_{\text{sat},i}(T)$ for all species, allowing us to obtain an initial guess for $y_i$ (for bubble-point calculations) or $x_i$ (for dew point-calculations). 
        
        With the saturation point of all the pure-species obtained, we also have the vapour and liquid volumes of each species. We can use these volumes to obtain initial guesses for the volumes of the mixture assuming ideal mixing:
        \begin{equation}
            V^\alpha = \sum_ix_i^\alpha V_i^\alpha\,.
            \label{eq:vol_0}
        \end{equation}
        
        There is one limitation to this method: when one of the components is supercritical. In this situation, we first obtain the critical point for all species and, if one component is indeed supercritical, we set:
        \begin{align}
            V_l &= V_c\,,\\
            V_v &= 1.2V_c\,,
        \end{align}
        
        and:
        \begin{equation}
            \tilde{p}_{\text{sat},i} = p(V_c,T)\,.
        \end{equation}
        where $\tilde{p}$ denotes that this is a fictitious saturation pressure as $T$ is greater than $T_c$ in this case. All of the above gives a rough estimate of what the `liquid-like' and `vapour-like' phases in the critical region would be for a particular species. These initial guesses have proven to be sufficiently robust for most systems of interest (fringe cases such as highly-asymmetric mixtures may struggle).
        \item Azeotrope points: This method is compatible with only binary mixtures. Here, we re-use the same initial point as bubble and dew points but we assume that the composition of both components in the liquid and vapour phase is 0.5. 
        \item LLE points: For mixtures excluding highly size-asymmetric ones, the LLE envelope is generally symmetric. Thus, for a given $\mathbf{x}$, we assume the composition in the second liquid phase is $1-\mathbf{x}$. The initial guesses for the liquid volumes are then obtained using equation \ref{eq:vol_0} with the volumes being obtained from the liquid volume of the saturation point for each species. 
        \item VLLE points: In the current implementation, without custom initial guesses, \texttt{Clapeyron.jl} can solve only for binary heterogeneous azeotropes where the initial guess for the composition of the liquid phases is [0.25,0.75] and [0.75,0.25], respectively, and the initial guess for the composition of the vapour phase is [0.5,0.5]. The initial guess for the volumes is obtained using equation \ref{eq:vol_0} (with the same treatment in the case where a single component is supercritical).
    \end{itemize}
    Note that the above is applicable to only equations of state as, in this case, equation \ref{eq:Mod_Raoult} can be used to solve for bubble and dew points in the case of activity-coefficient models. LLE and VLLE points cannot yet be solved for in \texttt{Clapeyron.jl} for activity-coefficient models. 
    
    Nevertheless, whilst these initial guesses work well in the general, simple cases there are a few limitations, some of which have already been mentioned above. One slightly more-obvious limitation is, for the bubble / dew point, at a given temperature / pressure and composition, it is possible for two points to correspond to different dew / bubble points. In order to deal with these cases where the initial guesses are not sufficiently close to the true or desired solution, all of the corresponding functions can take custom initial guesses as an optional argument, similar to the saturation pressure function. This gives users greater control over the property estimation methods, in contrast to `black-box' packages such as gPROMS\citep{2020GPROMS} or ASPEN\citep{2016AspenPlus}.
    
    Like the saturation points of pure components, these initial guesses serve only to find the pressure at these conditions. To find the temperature, one needs to add an additional layer of iterations to solve for the temperature corresponding to a given pressure (in a similar fashion to equation \ref{eq:p_sat_T}).
    
    \begin{figure*}[h!]
\centering
  \begin{subfigure}[b]{0.49\textwidth}
    \includegraphics[width=1\textwidth]{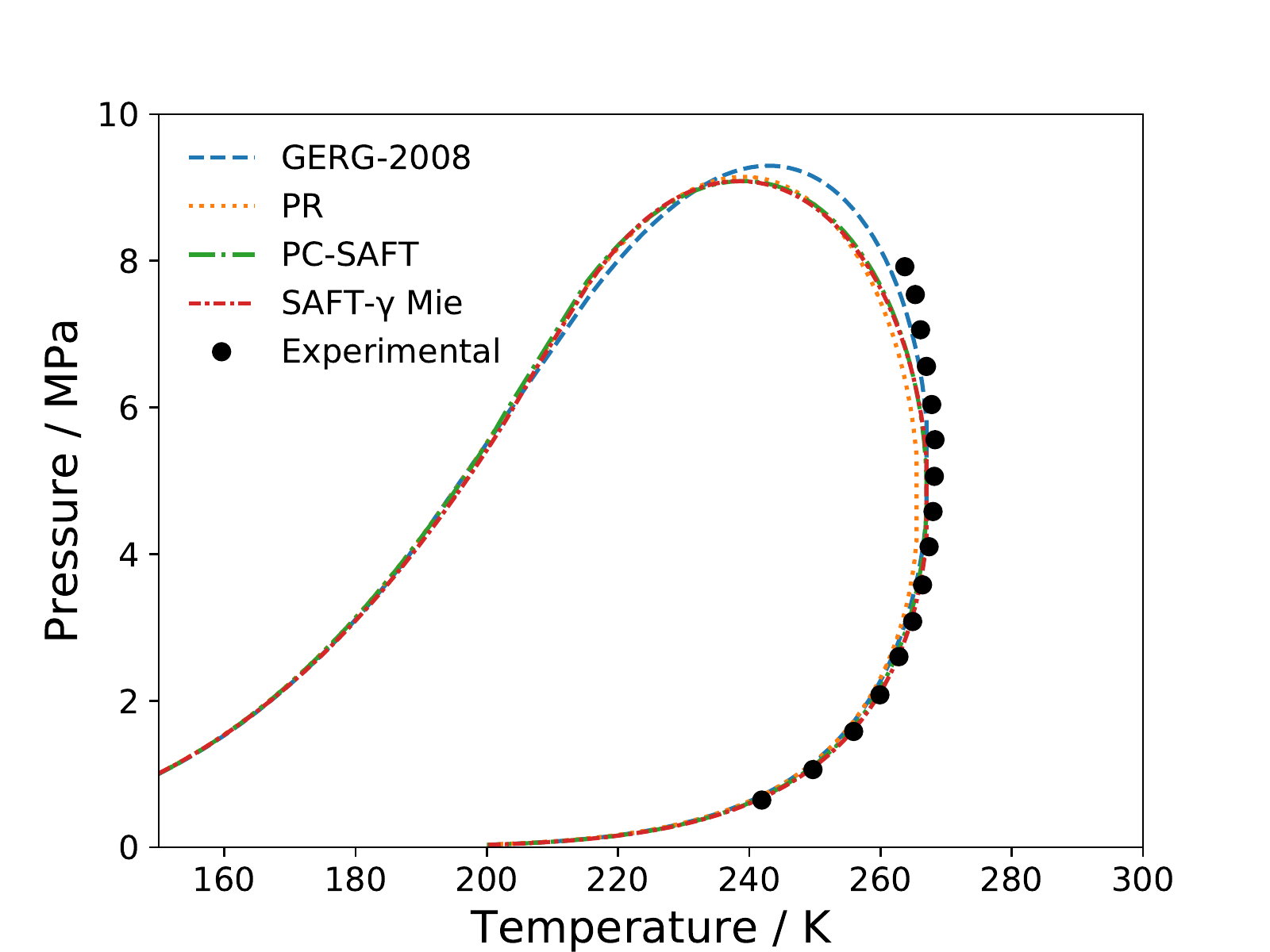}
    \caption{$pT$ isopleth for a four-component natural gas system.\citep{Avila2002ThermodynamicCorrelation}}
    \label{fig:4_component}
  \end{subfigure}
  \begin{subfigure}[b]{0.49\textwidth}
    \includegraphics[width=1\textwidth]{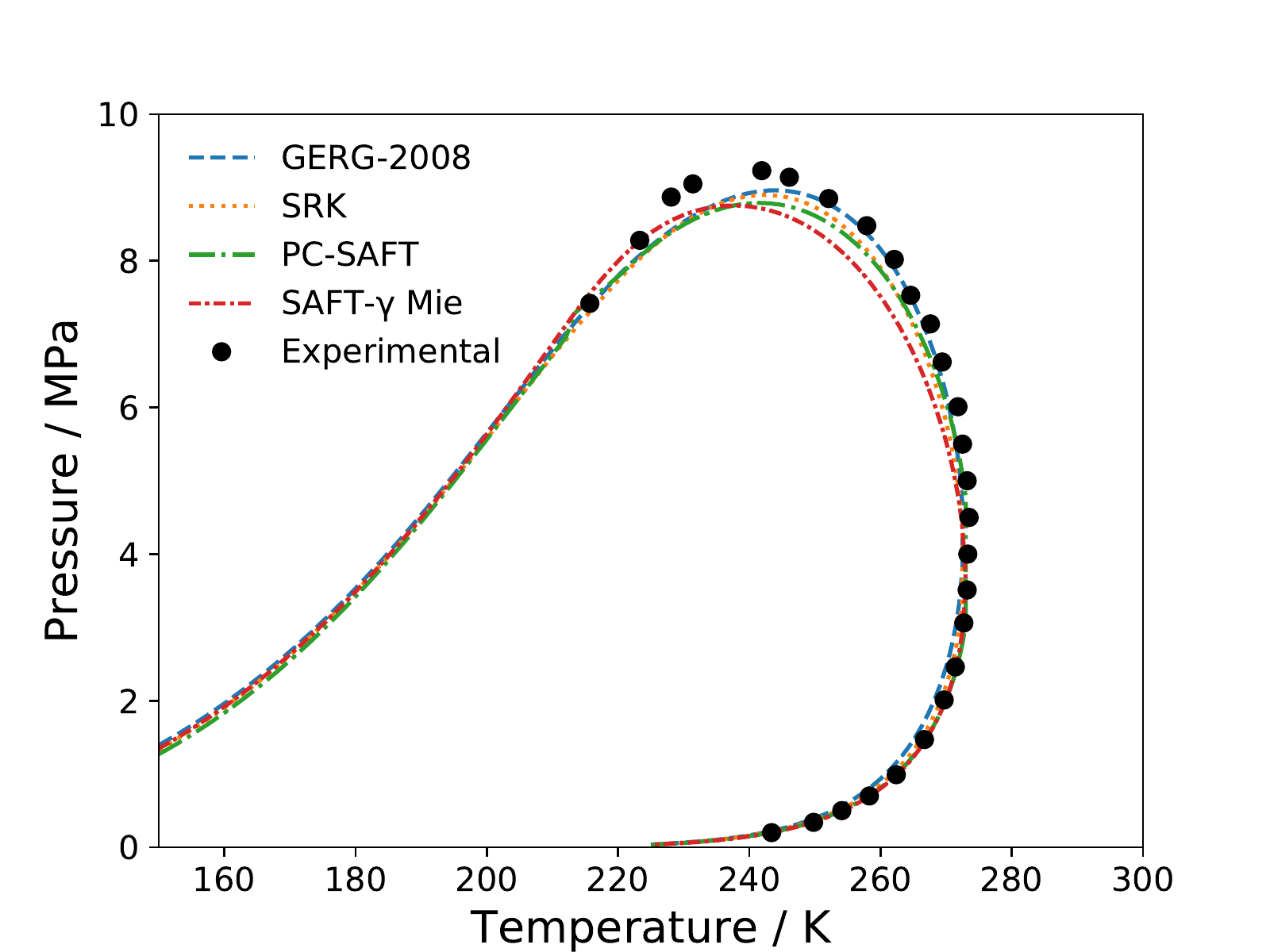}
    \caption{$pT$ isopleth for a 12-component natural gas system.\citep{Mrch2006MeasurementMixtures}}
    \label{fig:12_component}
  \end{subfigure}
  \caption{Examples of $pT$ isopleths obtained for large systems using various thermodynamic models within \texttt{Clapeyron.jl}.}
  \label{fig:multi_pT_isopleth}
\end{figure*}
    
    In combination, these methods can be used to generate a range of phase diagrams. Many of these have been shown in previous sections in the form of simple binary $pxy$ or $Txy$ diagrams (see figures \ref{fig:co2_co_alpha}, \ref{fig:meoh_benz_mix}, \ref{fig:water_co2_vle} and \ref{fig:act_water_e_H}). However, these methods can also be used for larger systems such as multi-component, $pT$ isopleths as shown in figures \ref{fig:4_component} (a four-component system) and \ref{fig:12_component} (a 12-component system) for multiple thermodynamic models (to our knowledge, this is the first time, in literature, SAFT-$\gamma$ Mie has been used for such large systems). Whilst these systems are liquid-natural gas systems with not-too-disimilar interactions between species, it is still impressive that \texttt{Clapeyron.jl} is able to generate these figures in seconds.
    
    \begin{figure*}[h!]
\centering
  \begin{subfigure}[b]{0.49\textwidth}
    \includegraphics[width=1\textwidth]{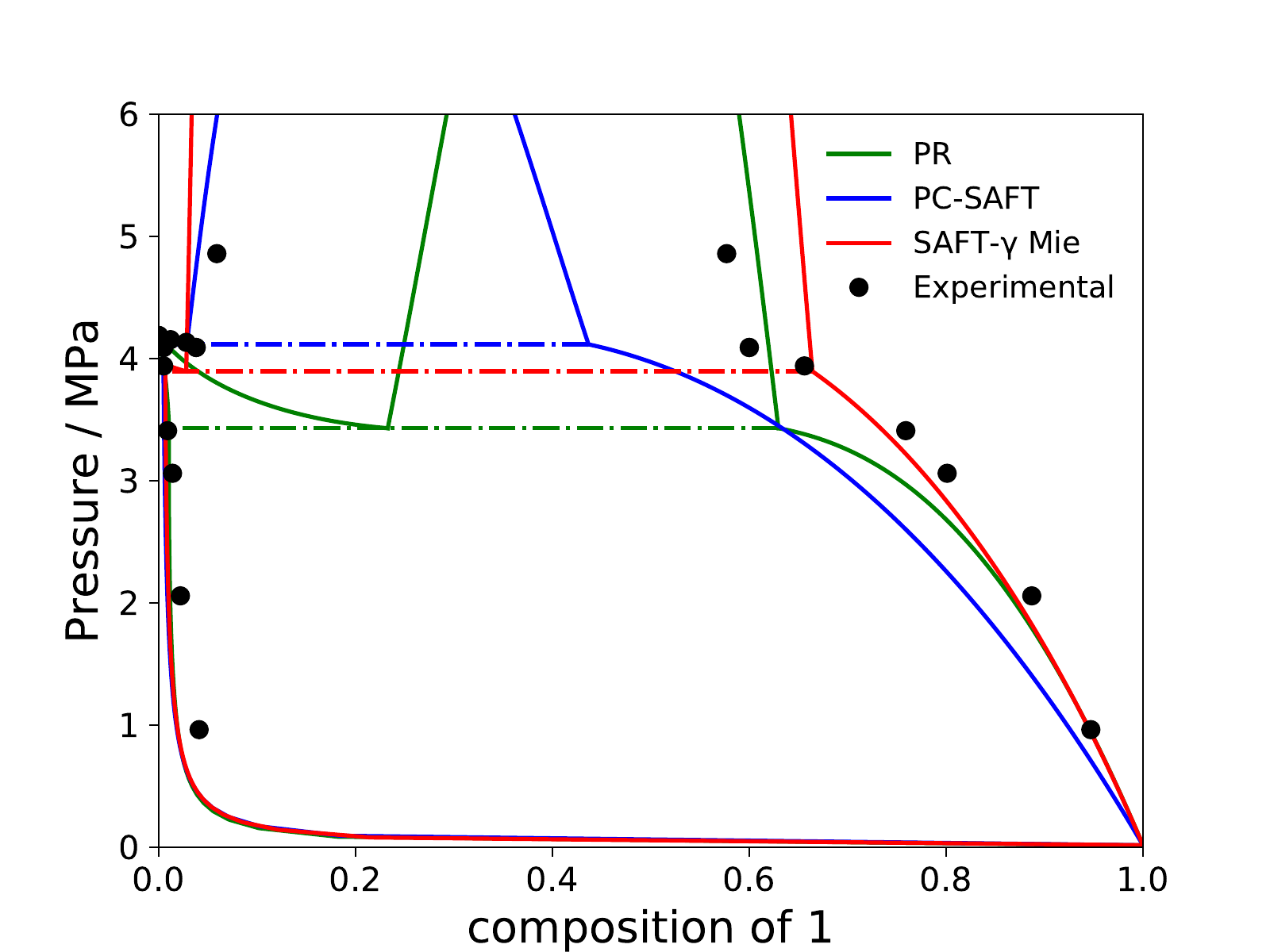}
    \caption{$pxy$ diagram for a methanol (1) + ethane (2) mixture at 298.15\,K.\citep{Ishihara1998PhaseK}}
    \label{fig:pxxy_meoh_eth}
  \end{subfigure}
  \begin{subfigure}[b]{0.49\textwidth}
    \includegraphics[width=1\textwidth]{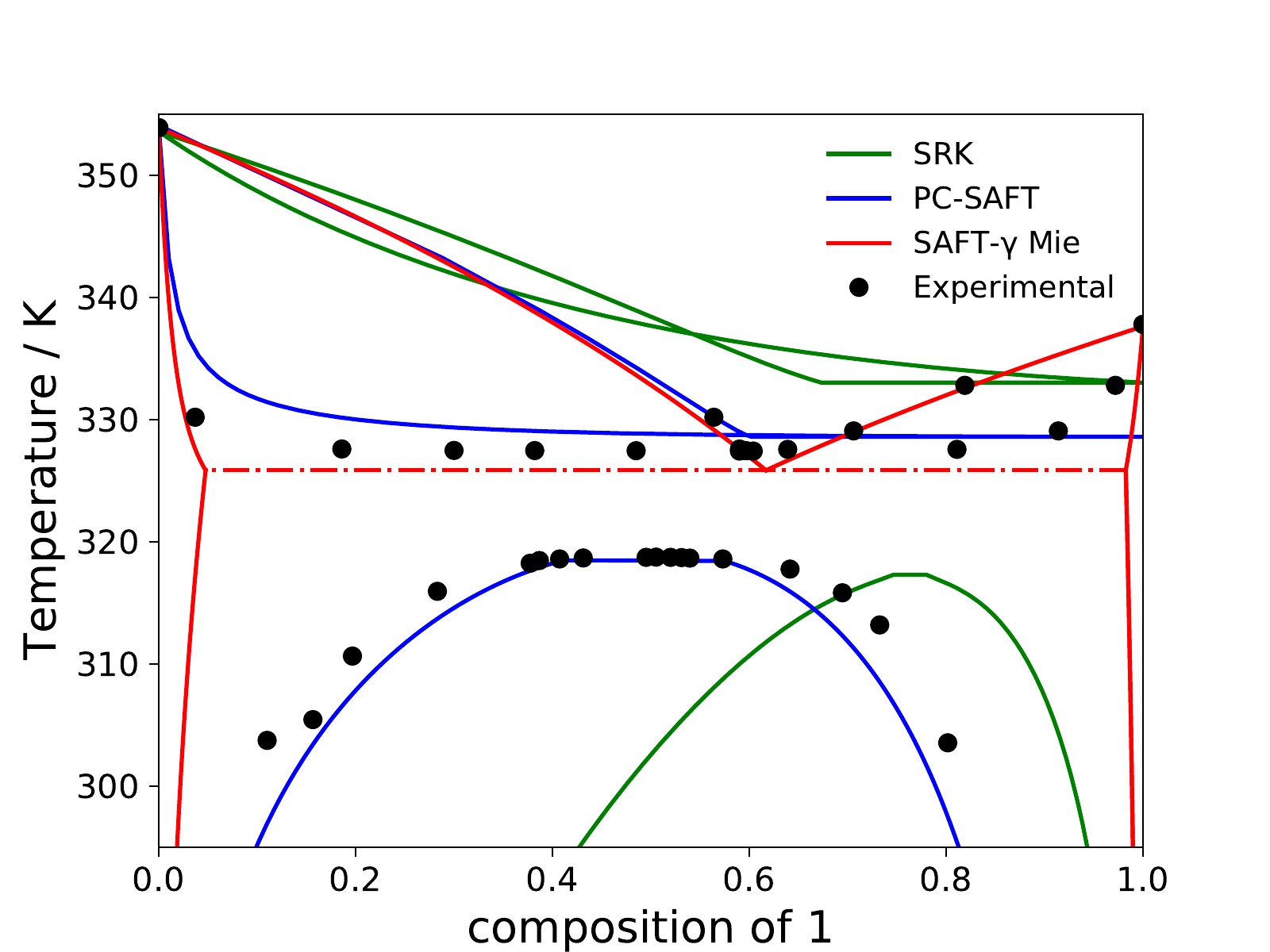}
    \caption{$Txy$ diagram for a methanol (1) + cyclohexane (2) mixture at 1.01\,bar.\citep{Jones1930TheAlcohol,Marinichev1965NoTitle}}
    \label{fig:Txxy_meoh_eth}
  \end{subfigure}
  \caption{Examples of VLLE phase diagrams for the methanol+ethane mixture obtained using various thermodynamic models within \texttt{Clapeyron.jl}.}
  \label{fig:VLLE_meoh_eth}
\end{figure*}
    
    Returning to binary systems, $pxxy$ and $Txxy$ diagrams are shown in figures \ref{fig:pxxy_meoh_eth} and \ref{fig:Txxy_meoh_eth}, with VLE, LLE and VLLE behaviour. This highlights that \texttt{Clapeyron.jl} is able model complex phase equilibria for multiple thermodynamic models. 
    \item Critical points: Similar to pure components, these are rarely supported in commercial packages. Even in packages in which they are supported, the critical point for mixtures is rarely solved for directly. This is primarily because of its complex definition\citep{Bell2017CalculationModels}:
    \begin{align}
        \mathcal{L}(V,T,z)=\det{\mathcal{H}(A(V,T,z)/(RT))} &= 0\,,\label{eq:crit1} \\
        \det{\begin{bmatrix}
\mathcal{H}^{N-1}(A(V,T,z)) \\
(\nabla \mathcal{L}(V,T,z))^T 
\end{bmatrix}} &= 0 \,,\label{eq:crit2}
    \end{align}
    where $\mathcal{H}$ denotes the Hessian operator, $\nabla$ denotes the gradient operator and the superscript $N-1$ denotes that we are only using the first $N-1$ rows of the Hessian. Without thinking about how to obtain analytical expressions for this system of equations, as one can clearly see, the derivatives involved in this system are very unusual and require that one have access up to the third order derivatives with respect to composition (without even considering the need for even higher order derivatives in root-finding algorithms). However, once again, \texttt{ForwardDiff.jl} comes to our aid and is able to provide us with all the derivatives that we need:
    \begin{minted}[breaklines,escapeinside=||,mathescape=true,  numbersep=1pt, gobble=2, frame=lines, fontsize=\small, framesep=2mm]{julia}
function Obj_crit_mix(model::EoSModel,F,z,V,T)
    f(x) = eos(model,V,T,x)
    H(x) = ForwardDiff.hessian(f,x)/R/T
    L(x) = det(H(x))
    dL(x) = ForwardDiff.gradient(L,x)
    M(x) = [H(x)[1:end-1,:];transpose(dL(x))]
    F[1] = L(z)
    F[2] = det(M(z))
    return F
end
    \end{minted}
    This definition for the critical point is true for both the critical point of VLE envelopes and Upper Critical Solution Temperatures (UCST). The only way to solve for one of these over the other is by using initial guesses:
    \begin{itemize}
        \item \texttt{crit\_mix}: This function will solve for the critical point of VLE envelopes at a given composition. The initial guesses for the critical temperature and volume is obtained using a geometric and arithmetic mean (scaled by their compositions) of the critical temperature and volume for the pure components, respectively. 
        \item \texttt{UCST\_mix}: This function will solve for the UCST at a given temperature (for a binary mixture) where initial guess for the composition is [0.5,0.5] and the initial guess for the volume is obtained using the strategy described in section \ref{sect:bulk} for the liquid volume at this composition.
    \end{itemize}
    The objective function for both of these properties is exactly the same and, like the previous properties, it is possible for users to provide their own initial guesses. Thus, it is technically possible for users to use the \texttt{crit\_mix} function to obtain the UCST at a given composition (the opposite will also be true) if they provide sufficiently accurate initial guesses. 
    
    One disadvantage of using this method when obtaining critical points for mixtures is that the computational costs will scale as, at minimum, $\mathcal{O}(N^5)$. Whilst this is still reasonable for binary mixtures, it does become very costly for larger mixtures. In these cases, it might be computational cheaper to simply solve for the point where the bubble and dew points are the same.
    
    \begin{figure*}[h!]
\centering
  \begin{subfigure}[b]{0.49\textwidth}
    \includegraphics[width=1\textwidth]{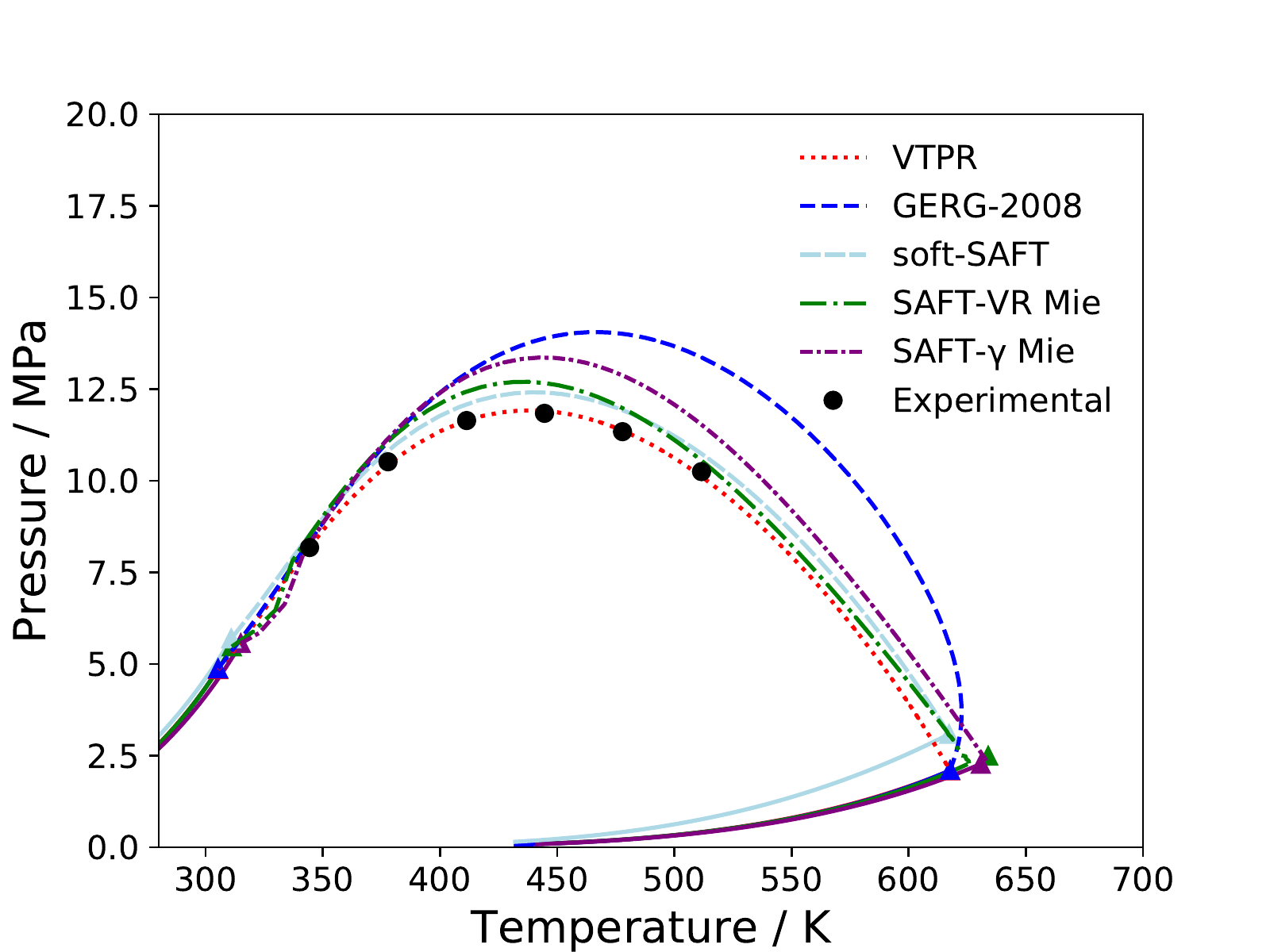}
    \caption{$pT$ projection of ethane+decane mixture.\citep{Reamer1962PhaseSystem.}}
    \label{fig:eth_dec_crit}
  \end{subfigure}
  \begin{subfigure}[b]{0.49\textwidth}
    \includegraphics[width=1\textwidth]{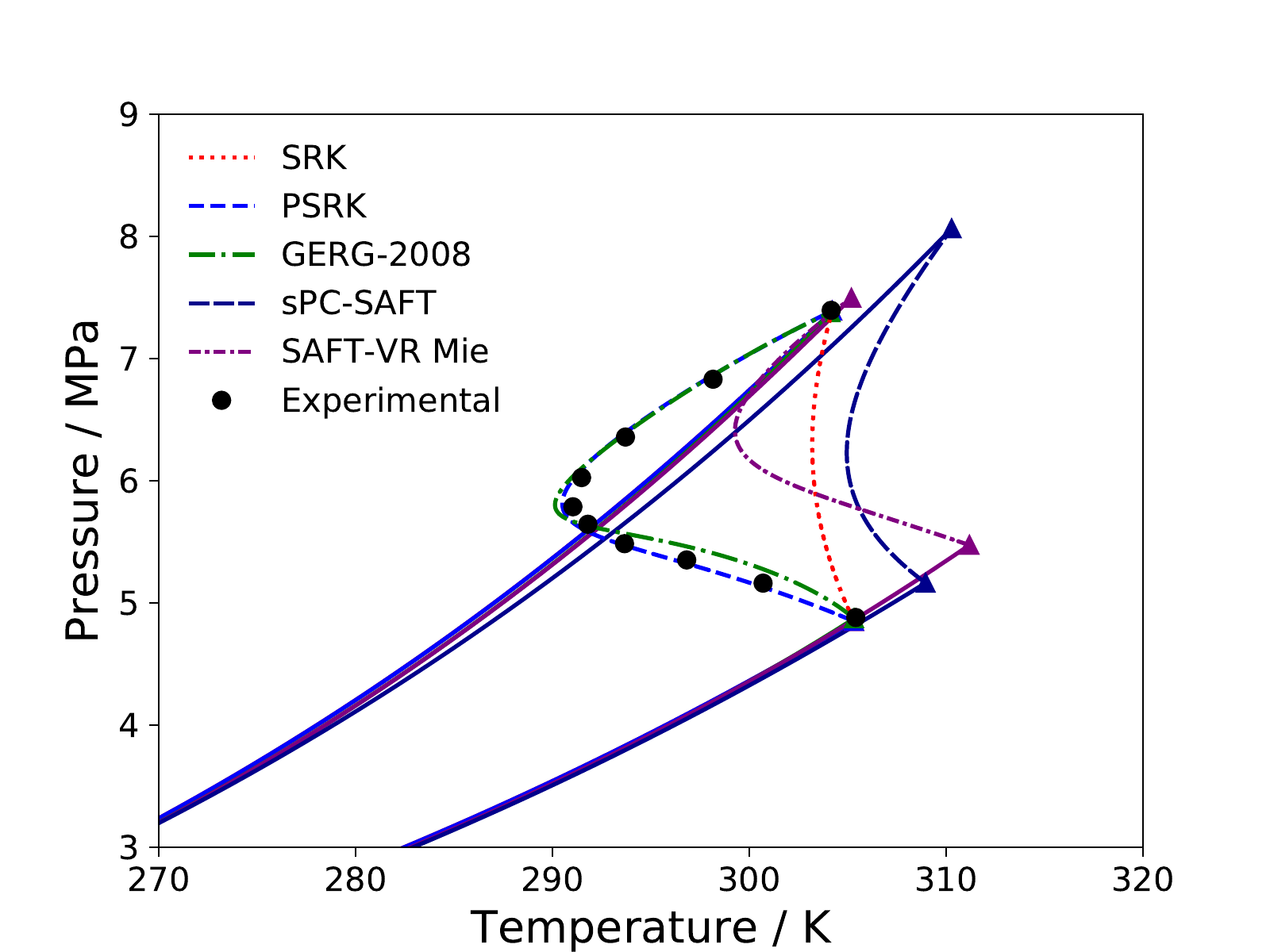}
    \caption{$pT$ projection of ethane+carbon dioxide mixture.\citep{Horstmann2000ExperimentalModel}}
    \label{fig:eth_co2_crit}
  \end{subfigure}
  \caption{Examples of $pT$ projections obtained using various thermodynamic models within \texttt{Clapeyron.jl}.}
  \label{fig:multi_pT_projection}
\end{figure*}
    
    Nevertheless, the flexibility of this implementation is again demonstrated in figures \ref{fig:eth_dec_crit} and \ref{fig:eth_co2_crit} where $pT$ projections for ethane+decane and ethane+carbon dioxide are shown. One can see the shape of the critical curve of the mixture doesn't affect \texttt{Clapeyron.jl}'s ability to model such systems.
    \item End points: These are points defined by a critical point in equilibrium with another phase. As such, we re-use equations \ref{eq:crit1} and \ref{eq:crit2}, and add the additional conditions for phase equilibria between two phases. Unfortunately, this is only currently supported for binary mixtures in \texttt{Clapeyron.jl} and can only find UCEPs for type II mixtures. For initial guesses, it is assumed that the critical / liquid phase has a composition of [0.5,0.5] and the vapour phase has a composition of [0.75,0.25]. The volumes are obtained using the same method as for the bubble point. 
    
    For this property, initial guesses become very important and, unless they are close to the true solution, the root finder will struggle to converge. Nevertheless, like all other VLE and critical properties, users are still able to provide custom initial guesses, allowing them to improve the convergence of the method and obtaining all other end-points.
\end{itemize}

\begin{figure}[h]
\centering
\includegraphics[width=0.7\textwidth]{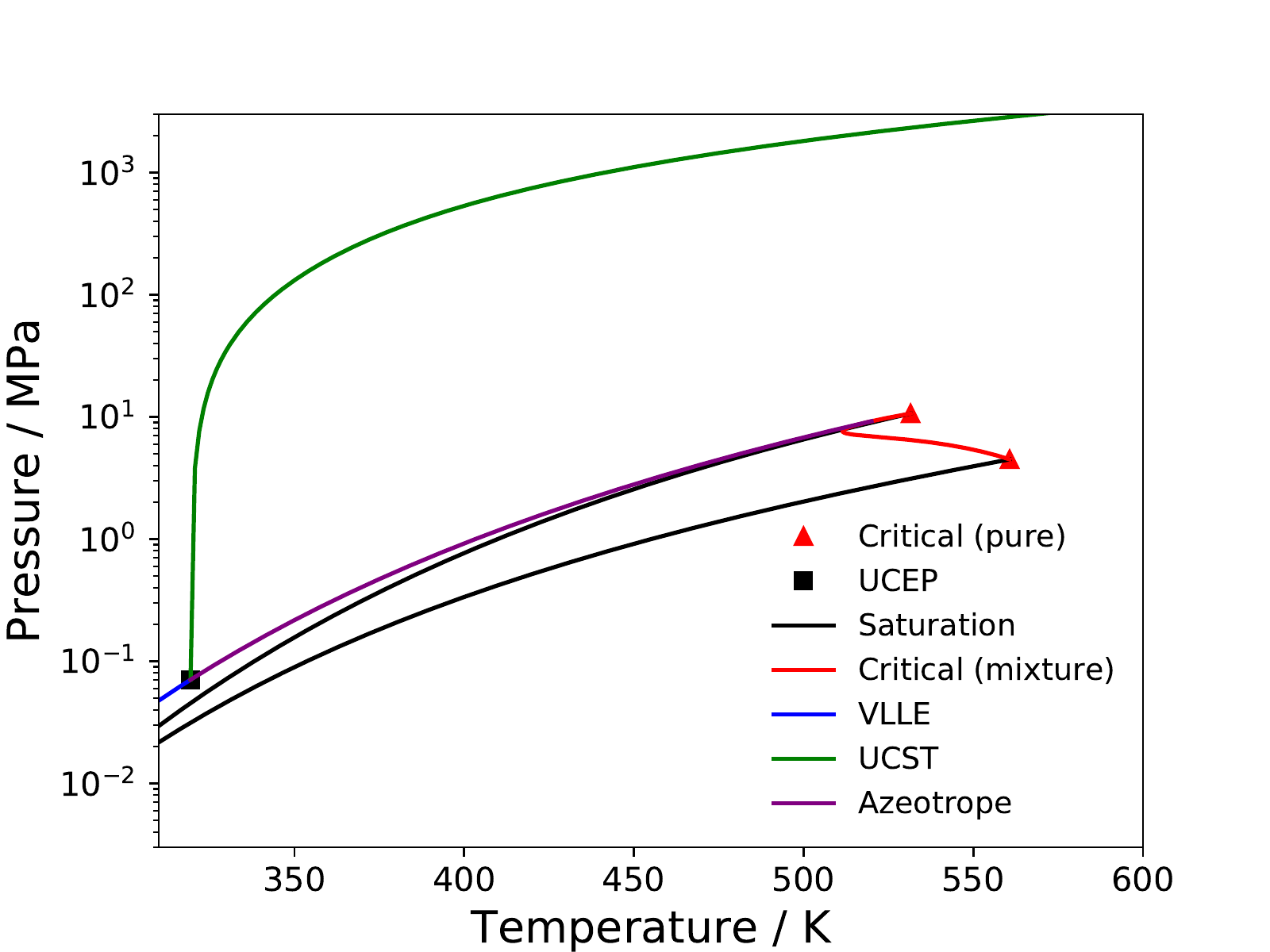}
\caption{$pT$ project of methanol+cyclohexane mixture, showing the saturation curves and critical points for each species, their critical, UCST, VLLE and azeotrope curves, along with the UCEP.}\label{fig:MeOH_Benz_pT}
\end{figure}

All of the VLE and critical properties listed in table \ref{tbl:props} are visually illustrated in figure \ref{fig:MeOH_Benz_pT}. This is a $pT$ projection of the methanol+benzene mixture, having the saturation curves for both components (ending at their critical points) connected by the critical curve. The azeotrope line ranges from the critical curve down to the UCEP where the UCST and VLLE curves all intersect. For visual clarity, this figure was made using only a single equation (PC-SAFT) but can be made using any other thermodynamic model with the supported parameters.

\section{Future work}
\label{sect:fw}
\texttt{Clapeyron.jl} is, by its very nature, a living entity in the sense that it is (and will remain) in continual development. Despite the many features currently available, there are still some desirable features currently in development: 
\begin{itemize}
    \item More thermodynamic models: Despite having an extensive list of thermodynamic models available in \texttt{Clapeyron.jl} (see table \ref{tbl:models}), there are still many left to implement from the literature. However, with its flexible framework, both the developers and users of \texttt{Clapeyron.jl} will be able to implement new models quickly (in the time taken to write this article, three new models were added). 
    
    Furthermore, whilst \texttt{Clapeyron.jl} can currently model a range of complex fluids, it can also be extended to other systems such as electrolytes. This is already in development on the \texttt{ElectrolyteModels} branch of the main repository and should be merged into the master branch soon.
    \item Flash routines: All the methods listed in section \ref{sect:methods} assumed prior knowledge of the phase equilibria. However, users may not always have this knowledge. Flash routines should solve for the `true' phase equilibrium at specified conditions (depending on the flash type). As such, they can be considered one of the most important tools in such a package. Development of these routines in \texttt{Clapeyron.jl} is already underway with two methods available: a rudementary Rachford--Rice algorithm\citep{Rachford1952ProcedureEquilibrium} and a differential evolution algorithm\citep{Bonilla-Petriciolet2017PhaseSearch}. It is our intention to soon implement the more-powerful, but more-complex flash routines developed by Michelsen and Mollerup\citep{Michelsen2007ThermodynamicAspects} and the HELD algorithm\citep{Pereira2012TheState}.
    \item Parameter estimation: Whilst users can plug-in existing parameters into \texttt{Clapeyron.jl}, obtaining new ones is equally as important. How to carry-out parameter estimation for thermodynamic models is a highly non-trivial issue\citep{Dufal2014,Dufal2015TheFluids}, involving degeneracies and being highly dependent on the dataset used. Development of a parameter estimation suit has begun on the \texttt{estimation} branch. Once fully implemented, \texttt{Clapeyron.jl} would be the first open-source package to provide all features normally provided by commercial software such as gPROMS\citep{2020GPROMS} and ASPEN\citep{2016AspenPlus} for thermodynamic equations of state.
\end{itemize}

Whilst the `permanent' development team of \texttt{Clapeyron.jl} is limited to the three authors of this article, it is our hope, with the gradual increase in usage of \texttt{Clapeyron.jl} that other individuals in the academic community will share their expertise and help develop \texttt{Clapeyron.jl}.
\section{Conclusion}
\label{sect:conc}
\texttt{Clapeyron.jl} is an open-source package for the implementation, application and development of thermodynamic models with a strong focus on transparency and extensibility. 

To the best of our knowledge, no other open-source package provides as wide a selection of thermodynamic models as \texttt{Clapeyron.jl}, which ranges from cubics equations of state, to SAFT-type equations of state, to activity-coefficient models, and various empirical Helmholtz models. Our implementation of cubic equations of state allows for a degree of customisability previously unseen in thermodynamic packages, giving users a simple way to modify their models straight from the front end. This level of flexibility is extended to the CPA equation of state, a unique aspect of our implementation. Whilst not as modular as cubics, a particularly wide range of SAFT-type equations has also been provided, all of which can be easily extended to provide the large variety of variants available in literature.

The property-estimation methods provided in \texttt{Clapeyron.jl} are also extensive, supporting most properties in which typical users might be interested. In this aspect particularly, automatic differentiation (through \texttt{ForwardDiff.jl}) has become a vital tool, not only allowing for easy computation of bulk properties, but also enabling \texttt{Clapeyron.jl} to support other properties, particularly critical properties of single and multi-component mixtures, that even commercial packages are unable to support. Whilst these methods may not be as robust as their commercial counterparts, users have been provided enough freedom to have complete control over the solvers and methods to compensate for this.

Unlike other open-source packages for thermodynamic models, \texttt{Clapeyron.jl} is readily and easily extensible with its multiple dispatch paradigm, as is the nature of packages written in Julia. Users can introduce their own parameters using simple-to-read-and-write CSV files, can implement variants of available models or entirely new ones and, develop or modify property estimation methods.

Finally, multiple features such as new thermodynamic models (including electrolyte models), flash routines and parameter-estimation tools are currently in development. With the open-source nature of \texttt{Clapeyron.jl}, the development of new features is not limited to what the `permanent' developers of the package are focused on. Other members of the community can share their expertise, develop their own features and integrate them with the package for the rest of the community to use.

As Clapeyron united the laws of Boyle, Charles, Avogadro and Gay-Lussac in the ideal gas law, leading to the development of all other thermodynamic models, we hope that \texttt{Clapeyron.jl} will serve as a unifying framework for all thermodynamic models and a platform for the development of subsequent models.

\section{Software Availability}
\label{sect:softav}
\texttt{Clapeyron.jl} is available through GitHub (https://github.com/ypaul21/Clapeyron.jl) and can be installed using:
\begin{minted}[breaklines,escapeinside=||,mathescape=true,  numbersep=1pt, gobble=2, frame=lines, fontsize=\small, framesep=2mm]{julia}
using Pkg
Pkg.add("Clapeyron")
\end{minted}
The notebook used to generate all the figures in this article can be found in the supplementary information and on the GitHub repository (https://github.com/ypaul21/Clapeyron.jl/examples). 

\section{Supplementary Information}
The following documents are provided:
\begin{itemize}
    \item \textbf{SI.pdf}: This document contains additional information on the structure of the databases used in \texttt{Clapeyron.jl}.
    \item \textbf{user\_defined\_eos\_example.ipynb}: This notebook contains an example of how users can define their own equation of state. The example pertains to the work by \citet{WalkerAbState}. We also provide the csv files needed to run these notebooks (\textbf{Ab\_initio\_SAFT.csv} and \textbf{Ab\_initio\_SAFT\_molarmass.csv}).
    \item We also provide the notebooks needed to generate all the figures presented in this work: 
    \begin{itemize}
        \item \textbf{cubic\_eos.ipynb}
        \item \textbf{SAFT\_eos.ipynb}
        \item \textbf{activity\_models.ipynb}
        \item \textbf{ideal\_eos.ipynb}
        \item \textbf{mixing\_functions.ipynb}
        \item \textbf{pure\_vle\_properties.ipynb}
        \item \textbf{multi-component\_vle\_vlle\_lle\_crit.ipynb}
    \end{itemize}
    All of these are also available on the GitHub repository for \texttt{Clapeyron.jl}.
\end{itemize}

\section{Acknowledgements}
The authors would like to thank Ramesh Putalapattu, Alex Ames, Sakse Dalum and Thomas Moore for contributing to the package. 
Very special thanks go to Dr. Andrew Haslam who initially introduced PJW and HWY to thermodynamics and has supported this endeavour since its infancy, providing vital feedback throughout, including the production of this manuscript. 
% \graphicalabstract{example-image-1x1}{Please check the journal's author guildines for whether a graphical abstract, key points, new findings, or other items are required for display in the Table of Contents.}

\bibliography{main.bib}
\end{document}

\maketitle
\section{Database}
The organisation of parameters plays a central role in thermodynamic modelling, as they tend to involve not just a description of each species, but also how they interact with each other. When there is intermolecular association involved (such as in SAFT-type equations), the complexity is further increased.

The full representation of all these thermodynamic parameters are often cumbersome and tedious to work with. Most existing applications to the authors' knowledge serialise these parameters into JSON or XML files, which can be difficult to read and write using scripts, let alone manually. For example, just to store (some) of the parameters associated with a single group in an XML file:
\begin{minted}{xml}
<groups>  
    <group name="CH3">
        <mw>15.0345</mw>
        <numberOfSegments>1</numberOfSegments>
        <shapeFactor>0.57255</shapeFactor>
        <selfInteraction>
                <dispersion>
                   <sigma>4.0772E-10</sigma>
                   <epsilon>256.77</epsilon>
                </dispersion>
        </selfInteraction>
    </group>
</groups>
\end{minted}

Alternatively, graphical user interface (GUI) are also used (such as in ASPEN) which, while accessible, quickly become impractical as the database grows in size.

One of the main design objectives for \texttt{Clapeyron.jl} is to make the databases as easy as possible to both read and write, even for less-experienced programmers.

\texttt{Clapeyron.jl} stores parameters in structs that are subtypes of \texttt{ClapeyronParam}. There are three main types of parameters, and we will use a parameter from each type from the PC-SAFT model for the water + carbon dioxide system to illustrate:
\begin{itemize}
    \item \texttt{SingleParam}: These are like parameters for a single species or group. For example, the critical temperature ($T_{c,i}$), acentric factor ($\omega_i$), number of segments ($m_i$), etc.
    \begin{minted}{julia}
julia> model.params.segment
SingleParam{Float64}("m") with 2 components:
 "water" => 1.0656
 "carbon dioxide" => 2.0729
\end{minted}

    \item \texttt{PairParam}: These are like and unlike parameters for a pair of species or groups. For example, binary interaction parameters ($k_{ij}$), potential parameters ($\epsilon_{ij}$ and $\sigma_{ij}$), like potential parameters, etc. Note the pair of species or groups need not necessarily be different.
    \begin{minted}{julia}
julia> model.params.sigma
PairParam{PairParam{Float64}}["water", "carbon dioxide"]) with values:
2×2 Matrix{Float64}:
 3.0007e-10   2.89295e-10
 2.89295e-10  2.7852e-10
\end{minted}

    \item \texttt{AssocParam}: These are association parameters for association interactions between two sites on two species or groups. For example, association energy (\(\epsilon_{ij,ab}^{\textrm{assoc.}}\)), bonding volume (\(\kappa_{ij,ab}\)), etc.
    \begin{minted}{julia}
julia> model.params.bondvol
AssocParam{Float64}["water", "carbon dioxide"]) with values:
("water", "e") >=< ("water", "H"): 0.034868
\end{minted}

\end{itemize}

These types of parameters are sufficient for most use cases one might encounter in equations of state. In the literature, these parameters are usually presented in tables corresponding to like, unilke and association parameters \citep{Huang1990EquationMolecules,Huang1991EquationMixtures,Yakoumis1997Vapor-liquidState,Gross2001Perturbed-ChainMolecules,Gross2002ApplicationSystems,Horstmann2005PSRKComponents,Haslam2020ExpandingMixtures}. To keep things as simple as possible from the point of view of the user, the databases used in \texttt{Clapeyron.jl} have been set-up in exactly the same way using CSV files:
\begin{itemize}
    \item Like parameters (usually named \texttt{MODEL\_like.csv}):
    \begin{table}[h!]
    \caption{Database set up for like parameters in \texttt{Clapeyron.jl}}
    \centering
\begin{tabular}{|l|l|l|}
\hline
\multicolumn{3}{l}{Clapeyron Database File}      \\\hline
\multicolumn{3}{l}{{MODEL NAME} Like Parameters} \\\hline
species         & param1         & param2        \\\hline
water           & 1234           & 5.678 \\\hline
\end{tabular}
\end{table}
    \item Unlike parameters (usually named \texttt{MODEL\_unlike.csv}):
    \begin{table}[h!]
    \caption{Database set up for unlike parameters in \texttt{Clapeyron.jl}}
    \centering
\begin{tabular}{|l|l|l|}
\hline
\multicolumn{3}{l}{Clapeyron Database File}      \\\hline
\multicolumn{3}{l}{{MODEL NAME} Unlike Parameters} \\\hline
species1          & species2        & param        \\\hline
water             & methanol        & 0.910 \\\hline
\end{tabular}
\end{table}
    \item Association parameters (usually named \texttt{MODEL\_assoc.csv}):
    \begin{table}[h!]
    \caption{Database set up for association parameters in \texttt{Clapeyron.jl}}
    \label{tbl:unlike_dtb}
    \centering
\begin{tabular}{|l|l|l|l|l|}
\hline
\multicolumn{5}{l}{Clapeyron Database File}      \\\hline
\multicolumn{5}{l}{{MODEL NAME} Assoc Parameters} \\\hline
species1   & site1  & species2  & site2  & param  \\\hline
water      & H      & water     & e      & 1.234  \\\hline
water      & H      & methanol  & e      & 5.678 \\\hline
\end{tabular}
\end{table}
\end{itemize}

In the case of unlike and association parameters, it is assumed by default that these parameters are symmetric (\textit{i.e.} $\sigma_{ij}=\sigma_{ji}$ or $\epsilon_{ij,ab}^{\textrm{assoc.}}=\epsilon_{ji,ba}^{\textrm{assoc.}}$). For models where that contain asymmetric unlike parameters like UNIFAC, the symmetric behaviour can be toggled off.

On the topic of UNIFAC, this model has its binary interaction parameters shared among a set of groups (termed sub-groups). For example CH$_3$, CH$_2$, CH and C are all sub-groups of a main CH$_3$ group. Rather than simply repeating this information, of which UNIFAC already contains over 1000 parameters for every possible combination of subgroups, it is possible to associate each row in the database with multiple species by using a special delimiter ($\sim$|$\sim$ by default). For example:

\begin{table}[h!]
    \caption{Database set up for unlike UNIFAC parameters in \texttt{Clapeyron.jl}}
    \centering
\begin{tabular}{|l|l|l|}
\hline
\multicolumn{3}{l}{Clapeyron Database File}      \\\hline
\multicolumn{3}{l}{UNIFAC   Unlike Parameters} \\\hline
species1              & species2    & A        \\\hline
CH3$\sim$|$\sim$CH2$\sim$|$\sim$CH$\sim$|$\sim$C    & ACH$\sim$|$\sim$AC    & 114.2  \\\hline
\end{tabular}
\end{table}

The CSV is chosen because it can easily be manipulated in plain text, or using a spreadsheet program.

\texttt{Clapeyron.jl} comes with a large selection of parameters for various models that are found in the literature. Alternatively, users will be able to, use their own parameters in \texttt{Clapeyron.jl} simply by including an optional argument (\texttt{userlocations}) in the function call to specify the location of the user-provided CSV files when creating the model object.
\begin{minted}{julia}
julia> model = PCSAFT(["my_molecule1","my_molecule2"];
userlocations=["my_parameters/"])
PCSAFT{BasicIdeal} with 2 components:
 "my_molecule1"
 "my_molecule2"
Contains parameters: Mw, segment, sigma, epsilon, epsilon_assoc, bondvol
\end{minted}

\texttt{Clapeyron.jl} is designed to be very flexible in terms of data types what can be stored within its databases. Parameters can be integers, floats, bools and strings, and the types are automatically inferred when the CSV file is read. It is also possible to store arrays of parameters (such as sigma profiles in the case of COSMO-SAC\citep{Bell2020ACOSMO-SAC}) by storing the array within a single cell.
\bibliographystyle{unsrtnat}
\bibliography{references.bib}